\documentclass[preprint,12pt]{elsarticle}

\usepackage{graphicx}
\usepackage{mathtools}
\usepackage{bm}
\usepackage{siunitx}
\usepackage{amsthm}
\usepackage{caption}
\usepackage{subcaption}

\usepackage{comment}
\usepackage[usenames,dvipsnames,svgnames,table]{xcolor}
\usepackage[color,matrix,arrow]{xy}
\usepackage{soul}

\usepackage{lineno}

\journal{Journal of the Mechanics and Physics of Solids}

\begin{document}

\begin{frontmatter}


\title{Spherical harmonics method for computing the image stress due to a spherical void}



\author{Yifan Wang, Xiaohan Zhang, Wei Cai}

\address{Department of Mechanical Engineering, Stanford University, CA 94305-4040, USA}


\begin{abstract}
We develop an efficient numerical method for calculating the image stress field induced by spherical voids in materials, and applied the method to dislocation-void interactions.
%
The method is constructed based on a complete set of basis functions for the displacement potential of the elastic boundary value problem for a spherical hole, as well as the corresponding displacement, stress, and traction fields, all in terms of linear combinations of spherical harmonics.
Using the fast transformation between the real and spherical-harmonics spaces provided by the SHTOOLS package, the method is more efficient than other image stress solvers such as the finite-element method.
This method can be readily extended for solving elasticity problems involving inclusions and inhomogeneities, as well as contact between spheres.  The tools developed here can also be useful for fast solution of differential equations with spherical boundaries beyond elasticity.
\end{abstract}


\end{frontmatter}

\nocite{*}


\section{Introduction}
\label{sec:intro}
Fracture of ductile materials is mainly controlled by the nucleation,
growth and coalescence of the voids in the material
(\citeauthor{TVERGAARD1989void_growth_coalesce},
\citeyear{TVERGAARD1989void_growth_coalesce}).
In continuum models, the growth of micro-voids
(\SI{>10}{\micro\meter}) can be described by the plastic flow
of the material around a spherical hole 
(\citeauthor{Benzerga2016ductile_failure},
\citeyear{Benzerga2016ductile_failure}).
For very small voids (\SI{<100}{nm}), such as during the initial
state of void growth, the kinetics is controlled by dislocation loop
emitting from the void 
(i.e. material being punched out from the void)
(\citeauthor{LUBARDA2004dislocation_emission},
\citeyear{LUBARDA2004dislocation_emission}). 
{This} dislocation loop emission mechanism has been studied 
by atomistic simulations 
(\citeauthor{TRAIVIRATANA2008void_growth_atomistic},
\citeyear{TRAIVIRATANA2008void_growth_atomistic};
\citeauthor{POTIRNICHE2006void_growth_coalesce_MD}, 
\citeyear{POTIRNICHE2006void_growth_coalesce_MD}a,
\citeyear{POTIRNICHE2006lattice_orientation}b).
However, for the intermediate-size voids (\SI{\sim1}{\micro\meter}),
the growth kinetics is too large for atomistic models and 
too small to be adequately described by continuum models.
The growth kinetics of micron-sized voids is influenced by
not only the dislocations emitting from the void surface, 
but also the nearby dislocations interacting with the void surface
(\citeauthor{CHANG2015size_effect_3ddd},
\citeyear{CHANG2015size_effect_3ddd}), 
which usually happens in the length scale of 
{\SIrange{1}{10}{\micro\meter}}, 
greatly exceeding the length scale of atomistic simulations.
At this {intermediate length-scale}, 
dislocation dynamics (DD) simulation becomes a suitable tool 
to study the growth kinetics of voids.

DD simulation is a method to quantitatively link the macroscopic plasticity of crystalline materials with the underlying dislocation structures and interactions (\citeauthor{arsenlis2007strain_hardening}, \citeyear{arsenlis2007strain_hardening}). Under a wide range of temperature and strain rate conditions, plastic deformation is dominated by dislocation motion and interactions. By considering only the dislocation lines instead of keeping track of all the atomic positions, DD simulations can reach a length scale of \SI{\sim10}{\micro\meter} at a time scale of \SI{1}{\micro\second} (\citeauthor{Sills2016DD}, \citeyear{Sills2016DD}), which is comparable to the scale of interest for micron-sized void growth kinetics. This means that potentially, it is possible to model {\it all} the dynamical behavior of dislocations around the void, the void-dislocation interactions, and their influence to the void growth kinetics, hence filling the gap between the atomistic modeling of the very small voids and the continuous modeling of the large voids.

Over the last ten years, DD simulations have been used to understand the size effect of void-dislocation interactions in two dimensions 
(\citeauthor{SEGURADO2009DDD_size_effect}, \citeyear{SEGURADO2009DDD_size_effect}, \citeyear{SEGURADO2010DDD_lattice_orientation}) 
and in three dimensions (\citeauthor{MUNDAY2015PDL_DD}, \citeyear{MUNDAY2015PDL_DD}). However, all these works have been limited to small strain and low dislocation density.
%
In these works, the finite-element method (FEM) has been used for solving the image stress needed to satisfy the boundary condition associated with the interaction between dislocations and the free surface of the void. Since the image stress needs to be solved at every iteration of the DD simulation, the accuracy and efficiency of DD simulations for void-growth problems have been limited by the FEM image stress solver.

In this work, we develop a semi-analytical {image stress solver for spherical interfaces}.
%
{Our method provides a good description of dislocation-void interactions when the void shape remains spherical.  This is expected to be the case when the surface steps produced by dislocations intersecting the void are much smaller than the void dimension and when the stress triaxiality is sufficiently high, where diffusion is rapid enough so that the void can maintain its spherical shape while growing in volume.}
We find a complete set of basis functions for the solution of elasticity problem represented in terms of linear combination of spherical harmonic functions.
{With the help of the }
fast spherical harmonic transformation provided by SHTOOLS (\citeauthor{shtools}, \citeyear{shtools}), our method finds a superposition of the basis functions that satisfies an arbitrary traction boundary condition on a spherical interface. 
{Because finding the correct superposition to satisfy the boundary condition requires only the solution of a sparse linear system on the spherical surface, this method has higher accuracy and efficiency compared with other solvers such as FEM and 3D-FFT} (\citeauthor{lebensohn2013fft_void_growth}, \citeyear{lebensohn2013fft_void_growth}), {which require 3D meshing of the material volume, and the boundary element method (BEM), which requires the solution of a dense linear system.}
The method relies on the availability of the 
coefficients of the basis functions and their displacement, stress and traction fields in terms of the spherical harmonic {expansion}.
%
These coefficients are pre-computed using 
ShElastic toolbox developed {in this work}
, which is built on top of a set of symbolic manipulation tools (specifically multiplication and spatial derivatives) for functions expressed in the spherical harmonics space.
These symbolic tools are likely to be useful for solving other partial differential equations with spherical boundaries beyond elasticity.
%
%

This paper is organized as follows: 
In section~\ref{sec:image-stress}, we formulate the image stress problem using the dislocation-void interaction as an example.
In section~\ref{sec:method}, we develop the spherical harmonic solution for the image stress problem.
In section~\ref{sec:result}, we present numerical results of several classical elasticity problems, including the dislocation-void interaction, to validate the method.
Discussions on further generalizations of the method are given in section~\ref{sec:discussion}.
The appendices provide derivations needed for the development of the ShElastic toolbox.

\section{Image stress problem}
\label{sec:image-stress}
\begin{figure}[!ht]
\centering\includegraphics[width=0.99\linewidth]{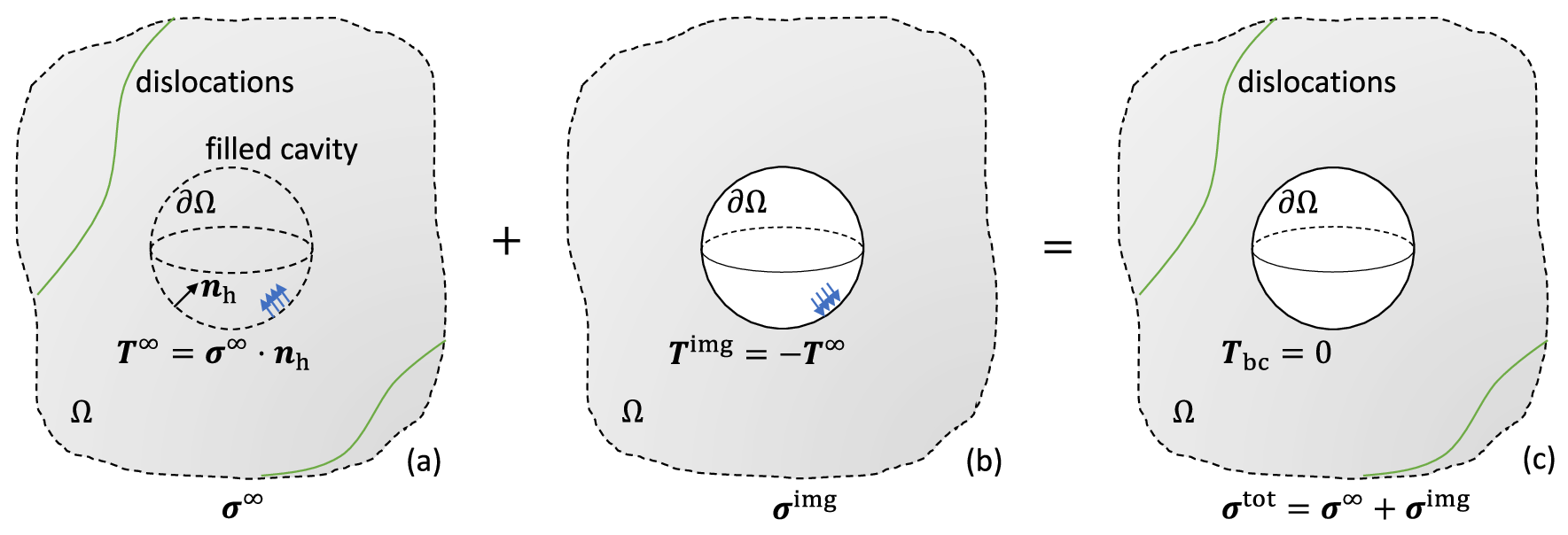}
\caption{Decomposition of the stress field in DD simulation. (a) Dislocations in an infinite medium without the spherical boundary $\partial\Omega$. (b) The image stress problem: an infinite medium containing the void with traction $T^{\rm img}$ on the void surface. (c) The original problem: an infinite medium containing dislocations and the void with zero traction on the void surface $\partial\Omega$ as a superposition of (a) and (b).}
\label{fig:decompose_problem}
\end{figure}

In DD simulations, the dislocation lines are discretized into straight segments connecting a set of nodes, which are the degrees of freedom (DOF) in the model (\citeauthor{arsenlis2007strain_hardening}, \citeyear{arsenlis2007strain_hardening}). In every time step of DD simulation, the most time consuming stage is evaluating the forces acting on the nodes. The Peach-Koehler force per unit length acting on the dislocation segment is calculated as:

\begin{equation}
\bm{f} = (\bm{\sigma}^{\rm tot}\cdot\bm{b})\times\xi
\label{eqn:Peach-Koehler}
\end{equation}

\noindent where $\bm{b}$ is the Burger's vector, and $\xi$ is the line direction vector. $\bm{\sigma}^{\rm tot}$ is the total stress field acting on the dislocation segment.
When the solid has a spherical (internal) surface $\partial\Omega$ that satisfies the traction-free boundary condition, the total stress field $\bm{\sigma}^{\rm tot}$ can be obtained as the superposition of two fields $\bm{\sigma}^{\infty}$ and $\bm{\sigma}^{\rm img}$ (\citeauthor{giessen1995discrete_dislocation}, \citeyear{giessen1995discrete_dislocation}), which are the solutions of two subproblems, as shown in figure~\ref{fig:decompose_problem} (a) and (b), respectively.

Let the stress induced by all the dislocations and external applied stress in an infinite elastic continuum (with no boundaries) be $\bm{\sigma}^{\infty}$, as shown in figure~\ref{fig:decompose_problem}(a), and let the corresponding displacement field be $\bm{u}^{\infty}$. Both $\bm{\sigma}^{\infty}$ and $\bm{u}^{\infty}$ of a given dislocation structure in an infinite medium can be obtained analytically (\citeauthor{hirth&lothe1982}, \citeyear{hirth&lothe1982}; \citeauthor{cai2006NSDD}, \citeyear{cai2006NSDD}). 
A traction field $\bm{T}^{\infty}$ on the spherical boundary $\partial \Omega$ is required if $\bm{\sigma}^{\infty}$ is applied to the domain $\Omega$, where $\bm{T}^{\infty} = \bm{\sigma}^{\infty} \cdot \bm{n}_{\rm h}$ and $\bm{n}_{\rm h}$ is the surface normal of the hole (i.e. $\bm{n}_{\rm h}$ points towards the center of the sphere). To satisfy the traction-free boundary condition in the original problem shown in figure~\ref{fig:decompose_problem}~(c), an image traction $\bm{T}^{\rm img}=-\bm{T}^{\infty}$ is introduced to cancel the $\bm{T}^{\infty}$ on the spherical surface $\partial\Omega$. The stress field in the continuum $\Omega$ induced by the image-traction boundary condition is called the image stress $\bm{\sigma}^{\rm img}$.  Here the {\it image stress problem} refers to the boundary-value problem on a spherical surface for solving $\bm{\sigma}^{\rm img}$ as shown in figure~\ref{fig:decompose_problem}(b). {The detailed definition of the boundary value problems of the three problems can be found in }\ref{adx:BVP}.

For certain boundary shapes, it is possible to use the symmetries of the geometry to derive semi-analytical solutions to reach both high efficiency and accuracy, such as {\citeauthor{WEINBERGER2007image_stress_cylinder} (\citeyear{WEINBERGER2007image_stress_cylinder})}, who developed a general method for solving the image stress problem on a cylindrical shape boundary based on fast Fourier transform (FFT).
For spherical-shaped boundaries, analytical solutions have only been developed for a number of specific problems. For example, \citeauthor{willis1972void_disl_interaction} (\citeyear{willis1972void_disl_interaction}) and \citeauthor{GAVAZZA1974screw_spherical_inclusion} (\citeyear{GAVAZZA1974screw_spherical_inclusion}) developed analytical solutions for the interaction between an infinite straight screw dislocation and a spherical void; and \citeauthor{sadraie2007spectral_inclusion}(\citeyear{sadraie2007spectral_inclusion}) {developed analytical solutions for the interaction between inclusions and cavities.}
For the void-growth problem which requires DD simulations with voids, it is of interest to develop a general image stress solver for spherical voids subjected to arbitrary image tractions (produced by curved dislocation lines) with higher accuracy and efficiency than other general methods for elasticity such as FEM.

In this paper, we develop such a semi-analytical method, which solves the image stress problem with any arbitrary traction on the boundary of a spherical hole in an infinite medium. 
Using this method, it is also easy to obtain the displacement field and the elastic energy, which are often of interest. In addition, the method can be readily generalized to solve elasticity problems with displacement (or mixed displacement-traction) boundary conditions on the void surface, as well as other types of elasticity problems involving spherical-shaped boundaries such as dislocation-inclusion interaction and contact between spheres.  The tools developed in this work may also find uses for solving other boundary-value problems beyond elasticity.

\section{General solution to the image stress problem with spherical boundary}
\label{sec:method}
For an isotropic linear elastic material without body force, the equilibrium condition can be written in terms of the displacement field $\bm{u}$ as:
\begin{equation}
\mu\nabla^2\bm{u}+(\lambda+\mu)\nabla(\nabla\cdot\bm{u})=0
\label{eqn:eq_cond}
\end{equation}

\noindent
where $\lambda$ and $\mu$ are Lam\'{e} constants.
We shall consider boundary condition as specified by the traction vector $\bm{T}^{\rm img}$ defined on the surface $\partial\Omega$, as shown in figure~\ref{fig:decompose_problem}(b).

In general, we solve the image stress problem in the following steps:
\begin{enumerate}
\item Based on the Papkovich-Neuber representation, we find a complete set of fundamental displacement solutions $\bm{u}^{(K)}$ that spans the solution space of equation~(\ref{eqn:eq_cond}). 
Using the ShElastic toolbox developed in this work (see~\ref{adx:detail_derivation}), the displacement field $\bm{u}^{(K)}$, and the corresponding stress field $\bm{\sigma}^{(K)}$ and traction $\bm{T}^{(K)}$ can be automatically generated and expressed as the superposition of spherical harmonic functions $Y_l^m$ (for definitions see~\ref{adx:spherical_harmonics}).

\item The user specified traction boundary condition $\bm{T}^{\rm img}(\theta,\varphi)$ is expanded in terms of the fundamental traction solutions $\bm{T}^{(K)}$:
\begin{equation}
\bm{T}^{\rm img}(\theta,\varphi)=\sum_{K}a_{K}\,\bm{T}^{(K)}(\theta,\varphi)
\label{eqn:traction_linear_system}
\end{equation}
\noindent
where $a_{K}$ are the expansion coefficients. This is accomplished by first expanding $\bm{T}^{\rm img}$ as a linear combination of $Y_l^m$
using SHTOOLS (\citeauthor{shtools}, \citeyear{shtools}), and then solve a set of linear equations. 

\item The image stress solution $\bm{\sigma}^{\rm img}$ can be obtained as a linear combination of the stress fields $\bm{\sigma}^{(K)}$:
\begin{eqnarray}
\bm{\sigma}^{\rm img}(r,\theta,\varphi)=\sum_{K}a_{K}\,\bm{\sigma}^{(K)}(r,\theta,\varphi)
\label{eqn:superposition_stress_mode}
\end{eqnarray}

\item Similarly, the displacement field $\bm{u}^{\rm img}$ in figure~\ref{fig:decompose_problem}(b) can be obtained as:
\begin{equation}
\bm{u}^{\rm img}(r,\theta,\varphi)=\sum_{K}a_{K}\,\bm{u}^{(K)}(r,\theta,\varphi)
\label{eqn:superposition_displacement_mode}
\end{equation}

\item Finally, after the displacement and the traction on the spherical interface is obtained, the elastic energy $E_{\rm el}^{\rm img} \equiv E_{\rm el}^{\rm tot} -  E_{\rm el}^\infty$ corresponding to the interaction between the dislocation and void can be calculated as:
\begin{equation}
E_{\rm el}^{\rm img}= \frac{1}{2} \int_{\partial\Omega} 
\bm{T}^{\infty}\cdot(\bm{u}^{\infty}+\bm{u}^{\rm img})\;dS
\label{eqn:image_elastic_energy}
\end{equation}
\noindent
where the integral is over the spherical surface. The proof of equation~(\ref{eqn:image_elastic_energy}) is provided in \ref{adx:image_energy}.
The integral can be easily evaluated in the spherical harmonics space as $4\pi$ times the complex inner product of the spherical harmonics coefficients
of $\bm{T}^{\infty}$ and $\bm{u}^{\infty}+\bm{u}^{\rm img}$. 
%
%

\end{enumerate}

\subsection{Represent fundamental solutions in spherical harmonic space}
\label{sec:decompose_modes}

In this section, we describe how to construct the fundamental displacement bases $\bm{u}^{(K)}$, the corresponding $\bm{\sigma}^{(K)}$, and $\bm{T}^{(K)}$ fields that span the solution space. In this work, we derive the solution based on the Papkovich-Neuber representation (\citeauthor{barber2009elasticity} 3rd ed., \citeyear{barber2009elasticity}), because the displacement potentials satisfy the Laplace's equation which have the general solution in terms of spherical harmonic functions. According to the Papkovich-Neuber solution, the following displacement field satisfies the equilibrium condition~(\ref{eqn:eq_cond}):

\begin{equation}
\bm{u}^{(K)}(r,\theta,\varphi) = \frac{1}{2\mu}
\left[-4(1-\nu)\bm{\psi}^{(K)}+\nabla\left(\bm{r\cdot\psi}^{(K)}\right)\right]
\label{eqn:displacement_mode}
\end{equation}

\noindent where the displacement potential $\bm{\psi}^{(K)}(r,\theta,\varphi)$ is a harmonic vector field, i.e. every component of $\bm{\psi}^{(K)}$ satisfies Laplace's equation $\nabla^2\psi_j^{(K)}=0,\;j=x,y,z$. 

The general solution of Laplace's equation $\nabla^2\omega(r,\theta,\varphi)=0$ in spherical coordinates can be represented as a linear combination of spherical harmonic functions times the radial function $(C^R_{lm}r^l+C^I_{lm}r^{-l-1})$:

\begin{equation}
\omega(r,\theta,\varphi)=
\sum_{l=0}^{\infty}\sum_{m=-l}^{+l}\left(C^R_{lm}r^l+\frac{C^I_{lm}}{r^{l+1}}\right)Y_l^m(\theta,\varphi)
\label{eqn:multipole_decomposition}
\end{equation}

\noindent
where $C^R_{lm}$ and $C^I_{lm}$ are constant coefficients, and $Y_l^m(\theta,\varphi)$ are the spherical harmonic functions with two angular indices $(l,m)$. Since there are multiple conventions in the literature, the precise definition of $Y_l^m$ used in this work is provided in \ref{adx:convention}. For the image stress of a spherical void in an infinite continuum, the $r^l$ terms can be eliminated since the solution should decay to zero as $r\rightarrow\infty$. Therefore, we choose the bases of the displacement potential as:

\begin{equation}
\bm{\psi}^{(K)}=\bm{\psi}^{(k,l,m)}(r,\theta,\varphi)=\hat{\bm{e}}_k\frac{1}{r^{l+1}}Y_l^m(\theta,\varphi)
\label{eqn:displacement_potential_mode}
\end{equation}

\noindent
where $\hat{\bm{e}}_k$ is the unit vector in the $k$-th direction $(k=x,y,z)$. 
Here we use index $(K)$ as a short-hand notation for 3-integer indices $(k,l,m)$. The completeness of this solution to the elasticity equation is proved in \ref{adx:harmonic_potential}. In this work, we only discuss the spherical void problem. However, adding the $r^l$ terms in the fundamental solution bases would lead to a solution applicable to linear elasticity problems in solid spheres and spherical inclusions.

Given a displacement potential basis $\bm{\psi}^{(K)}$, we can derive the displacement $\bm{u}^{(K)}(r,\theta,\varphi)$, the corresponding stress field $\bm{\sigma}^{(K)}(r,\theta,\varphi)$, and the traction vector $\bm{T}^{(K)}(\theta,\varphi)$ on the spherical surface $r=r_0$ in the following form:

\begin{eqnarray}
\label{eqn:displacement_solution}
u_j^{(K)}&=&u_j^{(k,l,m)}=\frac{1}{2\mu \, r^{l+1}} U^{(K)}_j(\theta,\varphi) \\
\label{eqn:stress_solution}
\sigma_{ij}^{(K)}&=&\sigma_{ij}^{(k,l,m)}= \frac{1}{r^{l+2}} S^{(K)}_{ij}(\theta,\varphi) \\
\label{eqn:traction_solution}
T_j^{(K)}&=&T_j^{(k,l,m)} = \frac{1}{r_0^{l+2}} F^{(K)}_j(\theta,\varphi)
\end{eqnarray}

\noindent 
where $U^{(K)}_j,S^{(K)}_{ij},F^{(K)}_j,\;(i,j=x,y,z)$ are the $r$-independent parts of displacement, stress, and traction respectively. 
Since any complex square-integrable function of $(\theta,\phi)$ can be expressed as a linear combination of $Y_l^m$ (equation~(\ref{eqn:sh_expand})), the orientation-dependence of the displacement $U^{(K)}_j(\theta, \varphi)$ can be represented as:
\begin{equation}
U_j^{(K)}(\theta,\varphi) = \sum_{l'=0}^{l'_{\max}}\sum_{m'=-l'}^{l'}\hat{U}^K_{jl'm'}Y_{l'}^{m'}(\theta,\varphi)
\label{eqn:SH_decompose_U}
\end{equation}
where $\hat{U}^K_{jl'm'}$ are the spherical harmonic expansion coefficients. Similarly, we can express the stress and traction in terms of spherical harmonics with expansion coefficients $\hat{S}^K_{ijl'm'}$, $\hat{F}^K_{jl'm'}$.
In addition, we use indices $(J)$ and $(I)$ as the shorthand notations for $(j,l',m')$ and $(i,j,l',m')$ respectively, so that the displacement, stress and traction fields can be represented by the expansion coefficients $\hat{U}^{K}_{J}$, $\hat{S}^{K}_{I}$, $\hat{F}^{K}_{J}$. 
These constant coefficients can be precomputed given the elastic constants $\mu$ and $\nu$, and can be used repeatedly for solving the image stress problem over multiple DD time-steps. The coefficients can be represented as a sparse matrix with $K$ as column index and $J$ (or $I$) as row index. 

The elementary operations needed for calculating the spherical harmonic expansion coefficients are the product of two functions and the spatial gradient of a function. For this purpose, we develop the ShElastic toolbox to obtain these coefficients analytically. 
Based on these two linear operations, the toolbox is able to calculate the spherical harmonic representation of displacement, stress, and traction based on the spherical harmonic bases of the displacement potential $\bm{\psi}^{(K)}$. In addition, the toolbox can be easily applied to other differential equations such as Love's solution for linear elasticity problems, or partial differential equations beyond elasticity (such as heat and electromagnetism). The detailed derivation of the toolbox is discussed in \ref{adx:detail_derivation}.

\subsection{Solve the image stress problem in spherical harmonic space}
\label{sec:solve_linear_system}

To simplify the notation, we express the radial coordinate $r$ in units of the void radius $r_0$, i.e. $r^*=r/r_0$. Then the boundary condition set by the image traction vector $\bm{T}^{\rm img}(\theta,\varphi)$ can be sampled on a $n\times2n$ regular mesh of the unit-sphere surface with $r^*=1$. The image stress and the displacement at any arbitrary point $(x,y,z)$ can be evaluated with scaled coordinates on the unit sphere, i.e. $\sigma_{ij}^{\rm img}(x,y,z)=\tilde{\sigma}_{ij}^{\rm img}(x^*,y^*,z^*),\;u_{j}^{\rm img}(x,y,z)=\tilde{u}_{j}^{\rm img}(x^*,y^*,z^*)$, where $x^*=x/r_0,\,y^*=y/r_0,\,z^*=z/r_0$.

We assume every component of traction $T^{\rm img}_j(\theta,\varphi),\;(j=x,y,z)$ is a square-integrable function, so that we can obtain the spherical harmonics coefficients through numerical spherical harmonic transform algorithms such as SHTOOLS with the maximum degree $l'_{\max}=n/2$:

\begin{equation}
T_j^{\rm img}(\theta,\varphi)=\sum_{l'=0}^{l'_{\max}}\sum_{m'=-l'}^{l'} \hat{T}^{\rm img}_{jl'm'} Y_{l'}^{m'}(\theta,\varphi)
\end{equation}

Again, we will use $(J)$ to represent the indices $(j,l',m')$. With the spherical harmonics coefficients $\hat{T}^{\rm img}_{J}$ and the coefficients matrix $\hat{F}^{K}_{J}$ of the fundamental solutions given in equation (\ref{eqn:traction_solution}), we can express the traction boundary condition (\ref{eqn:traction_linear_system}) in the following form:

\begin{equation}
\sum_{l'}\sum_{m'}\hat{T}^{\rm img}_{J}Y_{l'}^{m'}=\sum_{K}a_{K}\sum_{l'}\sum_{m'}\hat{F}^{K}_{J}Y_{l'}^{m'}
\end{equation}

\noindent which leads to,

\begin{equation}
\hat{T}^{\rm img}_{J}=\sum_{K}\hat{F}^{K}_{J}a_{K}
\label{eqn:simplified_linear_system}
\end{equation}

\noindent
where the coefficients $a_K$ can be solved from the set of linear equations (\ref{eqn:simplified_linear_system}).
Finally, the image stress solution can be reconstructed using the coefficients $a_{K}$:

\begin{equation}
\tilde{\sigma}_{ij}^{\rm img}(r^*,\theta,\varphi) = 
\sum_{l'=0}^{l'_{\max}}\sum_{m'=-l'}^{l'} \hat{\sigma}^{\rm img}_{ijl'm'}(r^*)Y_{l'}^{m'}(\theta, \varphi)
\label{eqn:image_stress_inverse_SHT}
\end{equation}

\noindent
where the coefficients are given by
\begin{equation}
\hat{\sigma}^{\rm img}_{ijl'm'}(r^*)
=\hat{\sigma}^{\rm img}_{I}(r^*)
=\sum_{K} \frac{a_{K}}{(r^*)^{l+2}}\hat{S}^{K}_{I}
\end{equation}
Therefore, the image stress field $\bm{\sigma}^{\rm img}$ at an arbitrary point $(r,\theta,\varphi)$ can be obtained by inverse spherical harmonic transform (equation (\ref{eqn:image_stress_inverse_SHT})) using SHTOOLS, i.e. by superposition of the spherical harmonics with coefficients $\hat{\sigma}^{\rm img}_{ijl'm'}(r^*)$ evaluated at $r^*=r/r_0$.

Similarly, the image displacement field can be obtained by superposing the fundamental solutions $\bm{u}^{(K)}$ as:

\begin{equation}
\tilde{u}_{j}^{\rm img}(r^*,\theta,\varphi) = 
\sum_{l'=0}^{l'_{\max}}\sum_{m'=-l'}^{l'} \hat{u}^{\rm img}_{jl'm'}(r^*)Y_{l'}^{m'}(\theta, \varphi)
\label{eqn:image_displacement_inverse_SHT}
\end{equation}

\noindent
where the coefficients are given by
\begin{equation}
\hat{u}^{\rm img}_{jl'm'}(r^*)
=\hat{u}^{\rm img}_{J}(r^*)
=\sum_{K} \frac{a_{K}}{(r^*)^{l+1}}\hat{U}^{K}_{J}
\end{equation}

\section{Numerical results}
\label{sec:result}
To verify the convergence and accuracy of the image stress method introduced above, we present three test cases for computing the image stress produced by (1) external loading, (2) an infinite straight screw dislocation, and (3) a prismatic dislocation loop, respectively. In the test case (3), we obtain the accurate numerical solution of the glide and climb forces on the prismatic dislocation loop size smaller and larger than the void size, which verifies the qualitative analysis by \citeauthor{willis1969dislocation_loop} (\citeyear{willis1969dislocation_loop}).

In addition, to illustrate the efficiency and accuracy of our method, we perform FEM calculations with ABAQUS (2017) running in serial mode. The FEM geometry takes the shape of a large cube containing a small spherical void and is discretized with second-order tetrahedral elements. The cube is 50 times larger than the void to minimize the external boundary effects of the cubic surfaces. The elements are refined around the spherical void and then gradually coarsened away from the void. The image stress values are obtained on Gauss points and then linearly interpolated along points of interest to compare with the spherical harmonics solution.

Both the spherical harmonic image stress solver (using pre-computed coefficients from ShElastic) and the FEM calculations are performed on a personal computer running {\tt Ubuntu16.04}. The spherical harmonics method is implemented in {\tt python3.6} and the sparse matrix solver is provided by {\tt numpy1.13} and {\tt scipy1.0}.

\subsection{Test case 1 -- Spherical void subjected to tensile load}
\label{sec:tensile_void}

\begin{figure}[!ht]
\centering
\begin{subfigure}[t]{2.6in}
	\centering
	\includegraphics[width=1.0\linewidth]{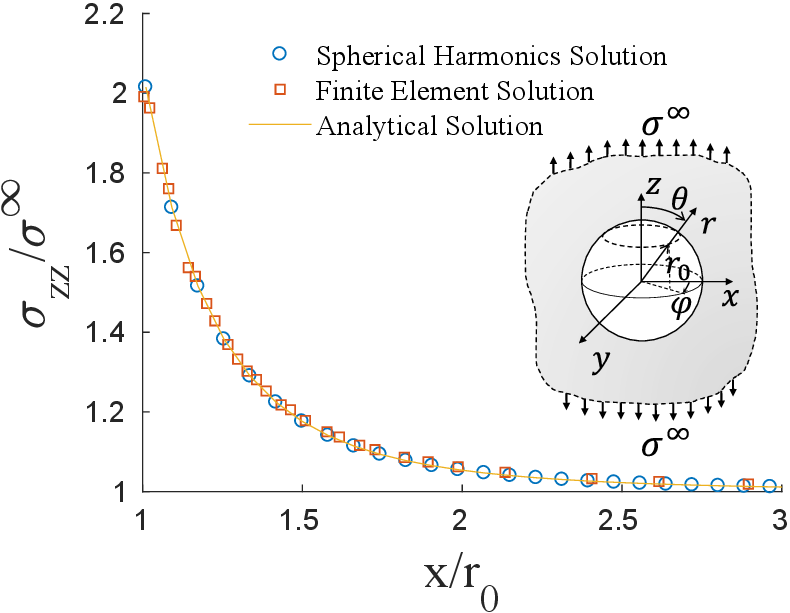}
    \caption{}\label{fig:tensile_hole_a}
\end{subfigure}\quad
\begin{subfigure}[t]{2.6in}
	\centering
	\includegraphics[width=1.0\linewidth]{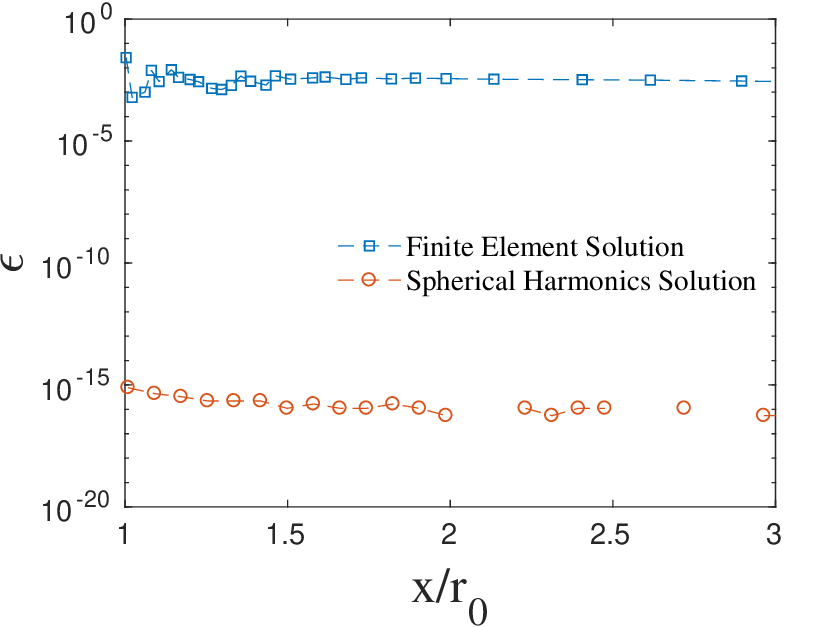}
    \caption{}\label{fig:tensile_hole_b}
\end{subfigure}
\caption{Spherical void in an infinite medium under tension. (a) Stress field $\sigma_{zz}(r)$ on the $x$-$y$ plane ($z = 0$) from spherical harmonics method, finite element method, and analytic solution.  The inset shows a sketch of the geometry. 
(b) Relative error of the numerical solutions against the analytic solution}
\label{fig:tensile_hole}
\end{figure}

We first consider the problem of a spherical void in an isotropic medium under uniaxial tension, as shown in figure~\ref{fig:tensile_hole}. The analytical solution of the stress field $\sigma_{zz}(r)$ on the $x$-$y$ plane ($z=0$) is given by \citeauthor{sadd2014elasticity} (\citeyear{sadd2014elasticity}):

\begin{equation}
\sigma_{zz} = \sigma^{\infty}\left[1+
	\frac{(4-5\nu)}{2(7-5\nu)}\left(\frac{r_0}{r}\right)^3+
    \frac{9}{2(7-5\nu)}\left(\frac{r_0}{r}\right)^5\right]
\end{equation}

\noindent
where $\sigma^{\infty}$ is the applied tensile stress; $r_0$ is the radius of the void; $\nu$ is the Poisson's ratio of the elastic medium.

To utilize our spherical harmonic solution, we first perform spherical harmonic decomposition of the image traction boundary condition on the void surface according to the equation (\ref{eqn:Y10_expression}):

\begin{equation}
\bm{T}^{\rm img}=-\bm{T}^\infty=-\hat{\bm{e}}_z\sigma^{\infty}\cos\theta=-\hat{\bm{e}}_z\sigma^{\infty}\sqrt{1/3}\,Y_1^0(\theta,\phi)
\end{equation}

\noindent The parameters in the calculation are taken as $\nu=1/3$, $\sigma^{\infty}=1$, $r_0=1$. The solution of the linear set of equations~(\ref{eqn:simplified_linear_system}), i.e. the coefficients $a_K$, is given in Table~(\ref{tab:case1-result}).

\begin{table}[ht]
\centering
\begin{tabular}{c c c}
\hline
$k$ \textbf{index} & $Y_l^m$ \textbf{index} & \textbf{Coefficient} $a_K = a_{klm}$\\
\hline
$x$ & $Y_1^{-1}$ & $-0.04523$ \\
    & $Y_1^1$    &  $0.04523$ \\
    & $Y_3^{-1}$ & $-0.02166$ \\
    & $Y_3^1$    &  $0.02166$ \\
\hline
$y$ & $Y_1^{-1}$ & $-0.04523\,i$ \\
    & $Y_1^1$    & $-0.04523\,i$ \\
    & $Y_3^{-1}$ & $-0.02166\,i$ \\
    & $Y_3^1$    & $-0.02166\,i$ \\
\hline
$z$ & $Y_1^0$    & $-0.2067$ \\
    & $Y_3^0$    & $0.03752$ \\
\hline
\end{tabular}
\caption{Non-zero coefficients in the spherical harmonic solution of image stress produced by tensile loading}
\label{tab:case1-result}
\end{table}

The solution converges with 10 modes ($l_{\max}=3$), and the modes with higher $l$ have zero coefficients. The total stress field can be evaluated as follows:

\[
\bm{\sigma}=(\hat{\bm{e}}_z\otimes\hat{\bm{e}}_z)\sigma^{\infty}+\bm{\sigma}^{\rm img}=(\hat{\bm{e}}_z\otimes\hat{\bm{e}}_z)\sigma^{\infty}+\sum_{K}a_{K}\bm{\sigma}^{(K)}
\]

We compare the stress on the $x$-$y$ plane with the analytical solution and FEM in figure~\ref{fig:tensile_hole}. FEM calculations of the image stress problem was performed with 84175 tetrahedral elements {and a total of 354300 degrees of freedom}. FEM has an maximum relative error of \SI{3}{\%} with a computation time of \SI{460}{s}. In comparison, the spherical harmonics image stress solver is able to reach the error of \num{3.5e-15} with the simulation time of \SI{0.026}{s}. Therefore, {our method is able to reach high accuracy with smaller sparse system, and thus higher efficiency compared with FEM.}

\subsection{Test case 2 -- Interaction between a void and a straight dislocation}
\label{sec:case_void_disl_interaction}

\begin{figure}[!ht]
\centering
\begin{subfigure}[t]{2.5in}
	\includegraphics[width=1.0\linewidth]{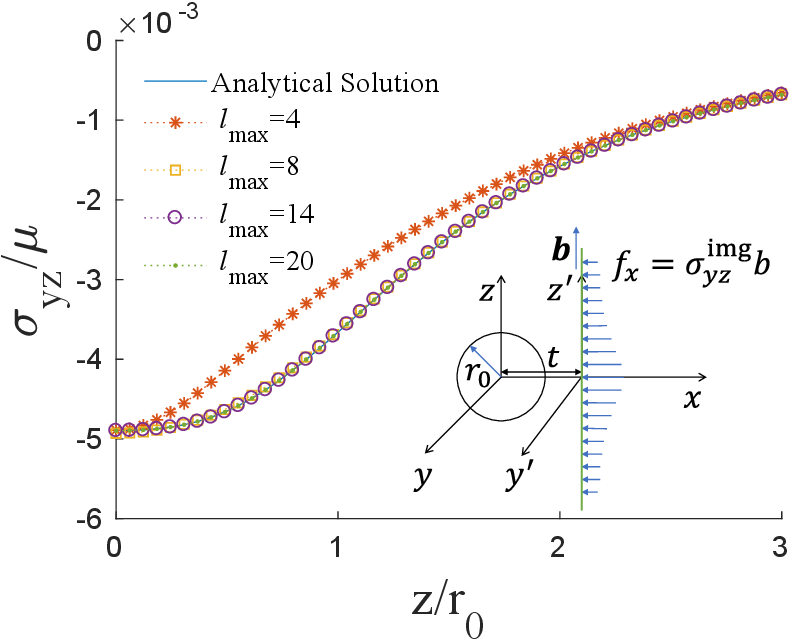}
	\caption{}\label{fig:result-void-disl-a}
\end{subfigure}\quad
\begin{subfigure}[t]{2.65in}
	\includegraphics[width=1.0\linewidth]{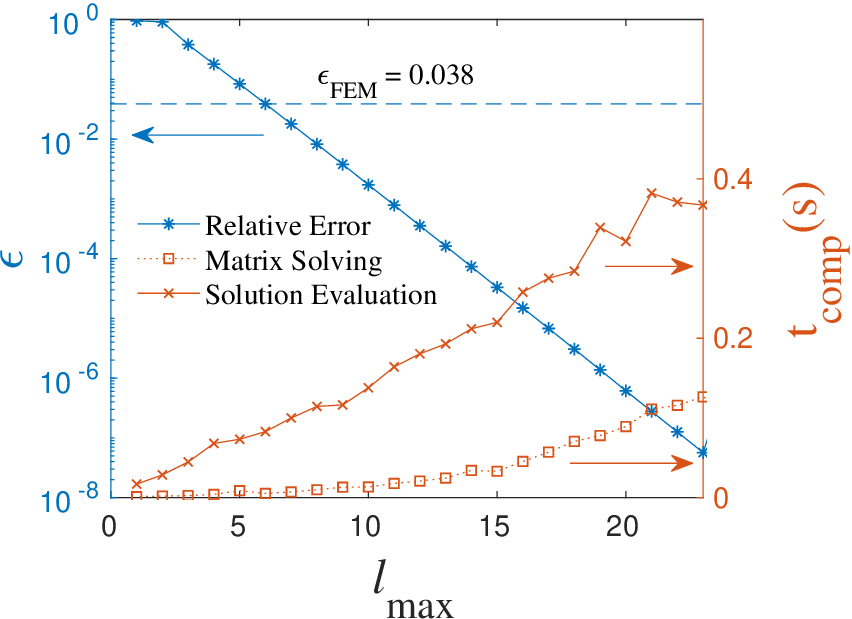}
	\caption{}\label{fig:result-void-disl-b}
\end{subfigure}
\caption{Comparison of spherical harmonics solution and analytical solution. (a) Image stress component $\sigma^{\rm img}_{yz}$ along the dislocation.  Inset shows a sketch of the geometry. (b) Relative error $\epsilon$ and computation time $t_{\rm comp}$ for different truncation values $l_{\max}$. The relative error of finite-element method $\epsilon_{\rm FEM}$ is marked as dashed line for comparison.}
\label{fig:result-void-disl}
\end{figure}

\citeauthor{GAVAZZA1974screw_spherical_inclusion} (\citeyear{GAVAZZA1974screw_spherical_inclusion}) and \citeauthor{willis1972void_disl_interaction} (\citeyear{willis1972void_disl_interaction}) developed analytical solutions independently, to the problem of the interaction between a spherical inclusion (with elastic constants $\lambda_2, \mu_2$) and an infinite long straight screw dislocation in an isotropic infinite medium (with elastic constants $\lambda_1, \mu_1$). These solutions reduce to the case of dislocation-void interaction in the limit of $\lambda_2,\mu_2 \to 0$. The geometry of the problem is shown in the inset of figure~\ref{fig:result-void-disl-a}. 

According to the symmetry of the problem, the Peach-Koehler force $\bm{f}$ given in equation (\ref{eqn:Peach-Koehler}) is only in the $x$-direction, and is a function of the void-dislocation distance $t$ and coordinate $z$ along the dislocation line: $f_x(t, z)=b\,\sigma^{\rm img}_{yz}(t,0,z)$, where $b$ is the magnitude of the Burger's vector and $\bm{\sigma}^{\rm img}$ is the image stress induced by the dislocations interacting with the void surface. The analytical solution of the image stress $\sigma^{\rm img}_{yz}(x,y,z)$ is given in \citeauthor{GAVAZZA1974screw_spherical_inclusion} (\citeyear{GAVAZZA1974screw_spherical_inclusion}).

In this test case, we set $\mu=\SI{52.5}{GPa}$, $\nu=1/3$, $t/r_0=1.5$, $r_0=\SI{1.25}{nm}$, and $b=\SI{0.25}{nm}$ (\citeauthor{STEINMETZ2013TWIP}, \citeyear{STEINMETZ2013TWIP}). The stress field of the dislocation in an infinite medium is:
\begin{equation*}
\sigma^{\infty}_{xz}=-\frac{\mu b}{2\pi}\frac{y}{(x-t)^2+y^2},\quad
\sigma^{\infty}_{yz}=\frac{\mu b}{2\pi}\frac{x-t}{(x-t)^2+y^2}
\end{equation*}
We truncate the solution to a finite series of spherical harmonics to achieve the required accuracy and efficiency. We truncate degree $l$ up to a maximum value $l_{\max}$, and for each $l$ we include all the $m$ values, $-l\leq m\leq l$, to preserve the spherical symmetry. 
Figure~\ref{fig:result-void-disl-a} shows the image shear stress $\sigma^{\rm img}_{yz}$ with different $l_{\rm max}$. At $l_{\max}=20$, the image stress computed by the numerical method agrees with the analytic solution very well. 
Figure~\ref{fig:result-void-disl-b} shows that the relative error compared to the analytical solution decreases exponentially with increasing $l_{\max}$. 

Figure~\ref{fig:result-void-disl-b} also shows that the matrix solving time increases quadratically with $l_{\max}$, and the solution evaluation time only increases linearly with $l_{\max}$. 
The quadratic scaling of the matrix solving time is consistent with the dimension (or rank) of the sparse matrix $\hat{F}_J^K$, which is on the order of $(l_{\max}+1)^2$. According to equation (\ref{eqn:superposition_stress_mode}), the coefficients $\hat{\sigma}_{ijl'm'}^{\rm img}(r^*)$ depends on the $r^{l+1}$, where $l$ is the second index of the fundamental solution $(K)=(k,l,m)$. The linear scaling of the solution evaluation time with $l_{\rm max}$ is due to the implementation in which the solutions with the same $l$-index are grouped and evaluated together.
Figure~\ref{fig:result-void-disl-b} shows that the method converges rapidly as $l_{\max}$ increases. For the case of $t/r_0=1.5$ with $l_{\rm max}=20$, our method already reaches a relative error of \num{e-6} using only \SI{0.38}{s} of computation time.

For comparison, we also performed FEM simulation of the void-dislocation interaction at $t/r_0=1.5$. In order to reach an acceptable accuracy, 84175 tetrahedral elements are used. The maximum relative error of the image stress calculated by FEM is marked as a blue dashed line in figure~\ref{fig:result-void-disl-b} for comparison with the spherical harmonics method. {The relative error of the FEM solution can be as high as \SI{3.8}{\%}, and the calculation takes \SI{387}{s}. Our spherical harmonics method outperforms the FEM method on both accuracy and efficiency by almost four orders of magnitude in the void-dislocation interaction calculation. We expect our method to achieve even higher efficiency with a {\tt C} or {\tt Fortran} implementation, which would significantly improve the efficiency of DD simulation of void growth problems.

\begin{figure}[!ht]
\centering
\begin{subfigure}[t]{2.6in}
	\includegraphics[width=1.0\linewidth]{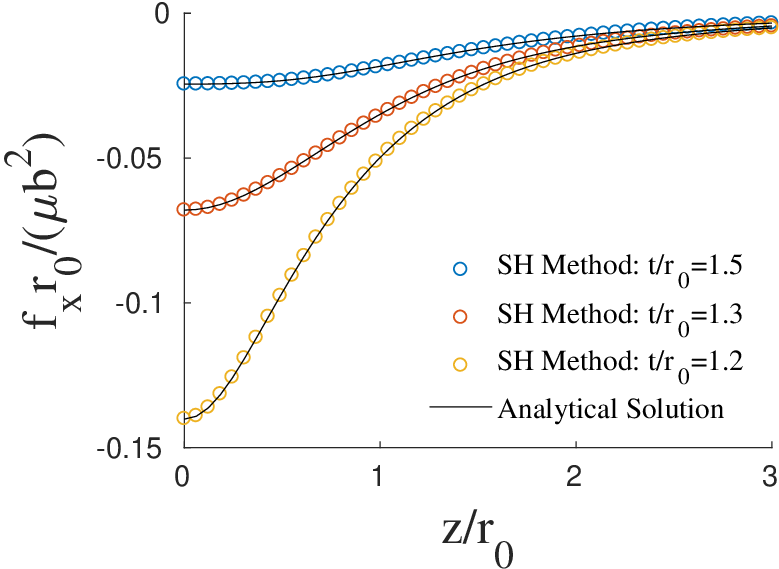}
	\caption{}\label{fig:result-void-disl-ts-a}
\end{subfigure}\quad
\begin{subfigure}[t]{2.6in}
	\includegraphics[width=1.0\linewidth]{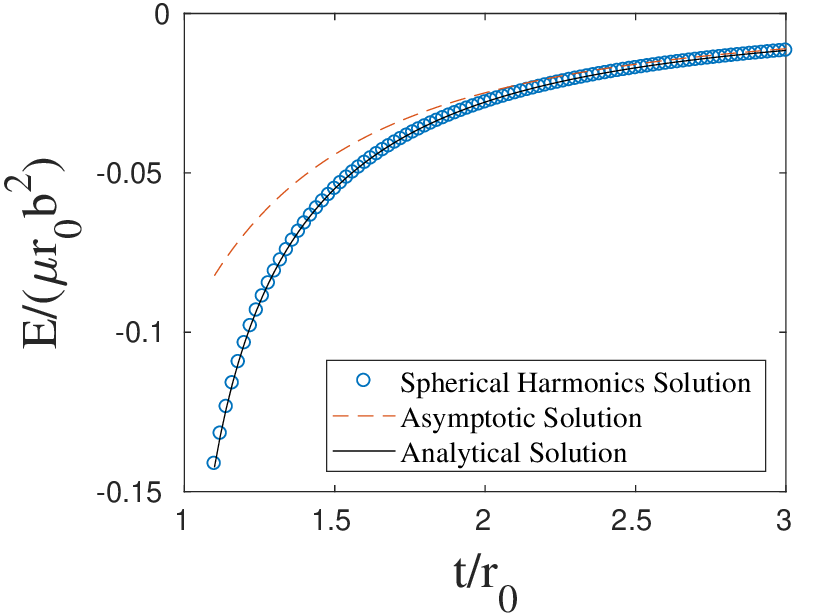}
	\caption{}\label{fig:result-void-disl-ts-b}
\end{subfigure}
\caption{Void-dislocation interaction along the dislocation with different stand-off distance $t$. (a) Interaction force $f_x(t, z)=b\,\sigma^{\rm img}_{yz}(t,0,z)$ along the dislocation. Stars: numerical solution from this work.  Lines: analytic expression from \citeauthor{GAVAZZA1974screw_spherical_inclusion}~(\citeyear{GAVAZZA1974screw_spherical_inclusion}). (b) Image elastic energy $E_{\rm el}^{\rm img}$ as a function of the stand-off distance $t$.  Circles: numerical solution from this work. Solid line: analytic expression from \citeauthor{willis1972void_disl_interaction}~(\citeyear{willis1972void_disl_interaction}).  Dashed line: asymptotic interaction energy given by equation~(\ref{eq:Eimgasympt}). }
\label{fig:result-void-disl-ts}
\end{figure}

Figure~\ref{fig:result-void-disl-ts}(a) plots the Peach-Koehler force from the image stress on the dislocation line. Very good agreement is observed between our numerical results and the analytic solution \citeauthor{GAVAZZA1974screw_spherical_inclusion}~(\citeyear{GAVAZZA1974screw_spherical_inclusion}).
To further validate our method, we calculate the image energy $E_{\rm el}^{\rm img}$, which is the interaction energy between the dislocation and the void, as a function of the stand-off distances $t$.
The analytic expression of $E_{\rm el}^{\rm img}$ can be obtained from the more general result in~\citeauthor{willis1972void_disl_interaction} (\citeyear{willis1972void_disl_interaction}). The leading term can be written in the explicit form:
\begin{equation}
 E^{\rm asy}(t) = -\frac{5\mu b^2\,r_0^3}{2\pi t^2}\frac{1-\nu}{7-5\nu}
 \label{eq:Eimgasympt}
\end{equation}
This is the dominant term at large $t$, and will be referred to as the asymptotic image energy.
The $1/t^2$ scaling of the asymptotic image energy is expected because the stress field of the dislocation scales as $1/t$ so that the elastic energy stored in a spherical volume of an infinite medium should scale as $1/t^2$ for very large $t$.

Figure~\ref{fig:result-void-disl-ts}(b) compares our numerical results for the image energy $E_{\rm el}^{\rm img}$ against the analytic expression.  Again, very good agreement is observed between the two; the maximum relative error is less than \num{0.8}\% at $t=1.1\,r_0$ with $l_{\max} = 20$.
While the asymptotic expression provides a good description for the dislocation-void interaction energy at large $t$, it significantly underestimates the magnitude of the interaction at small stand-off distances (e.g. by $\sim 12\%$ for $t < 1.5\,r_0$).

\subsection{Test case 3 -- Interaction between void and a prismatic dislocation loop}
\label{sec:case_void_PDL_interaction}

One of the key mechanisms of void growth is the interaction between the void and a nearby prismatic dislocation loop (PDL) (\citeauthor{LUBARDA2004dislocation_emission}, \citeyear{LUBARDA2004dislocation_emission}). Several analytical methods have been proposed to model this interaction and its relation to void growth, such as \citeauthor{AHN2006micromechanics_PDL_emission} (\citeyear{AHN2006micromechanics_PDL_emission}).
In this section, we couple our image stress solver to a DD simulation program, DDLab (\citeauthor{cai2006NSDD}, \citeyear{cai2006NSDD}), to demonstrate the capabilities of our approach.

The coupling between our solver and the DD simulation is achieved as follows. In every iteration, we let DDlab export the stress field of the dislocation in an infinite medium and calculate the traction force on the void surface. We then feed the negative traction force as the boundary condition to spherical harmonics solver to obtain the image stress solution $\bm{\sigma}^{\rm img}$. Finally, we evaluate the Peach-Koehler forces acting on the dislocation segments and add their contributions to the nodal forces calculated by DDlab, and let the DD simulation evolve.

We performed two DD simulations of a PDL in the neighborhood of the void, with the loop radius $\rho_0 = 0.75\,r_0$ (the loop is smaller than the void) and $\rho_0 = 1.2\,r_0$ (the loop is larger than the void). The elastic properties for the medium are $\mu = 52.5$ {GPa}, $\nu = 1/3$, same as in Section~\ref{sec:case_void_disl_interaction}.
We only consider the glide motion of the loop so that the loop radius remains constant in the simulation.
In addition, we do not consider the applied stress $\bm{\sigma}^{\rm app}$ and only consider the effect of the image stress $\bm{\sigma}^{\rm img}$ on the dislocation loop.

\begin{figure}[!ht]
\centering
\begin{subfigure}[t]{2.6in}
	\includegraphics[width=1.0\linewidth]{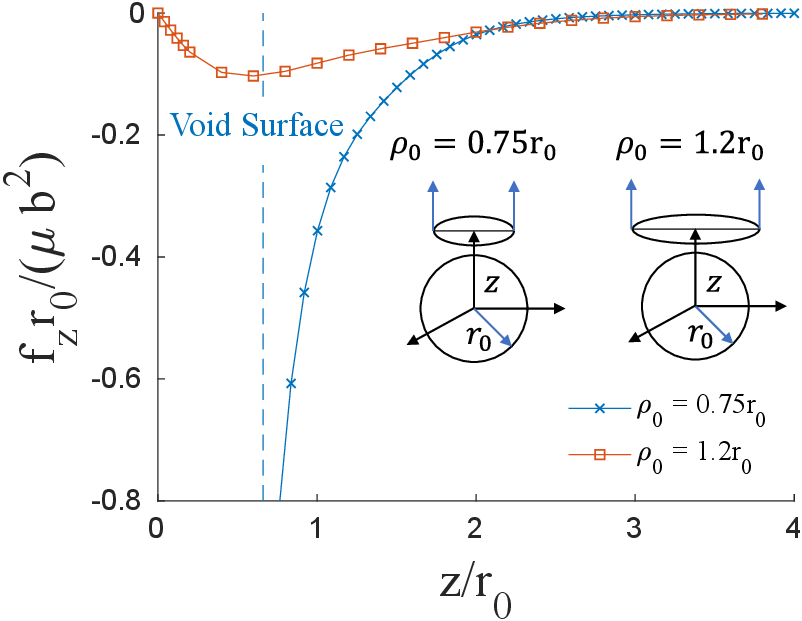}
	\caption{}
    \label{fig:result-void-pdl-a}
\end{subfigure}\quad
\begin{subfigure}[t]{2.6in}
	\includegraphics[width=1.0\linewidth]{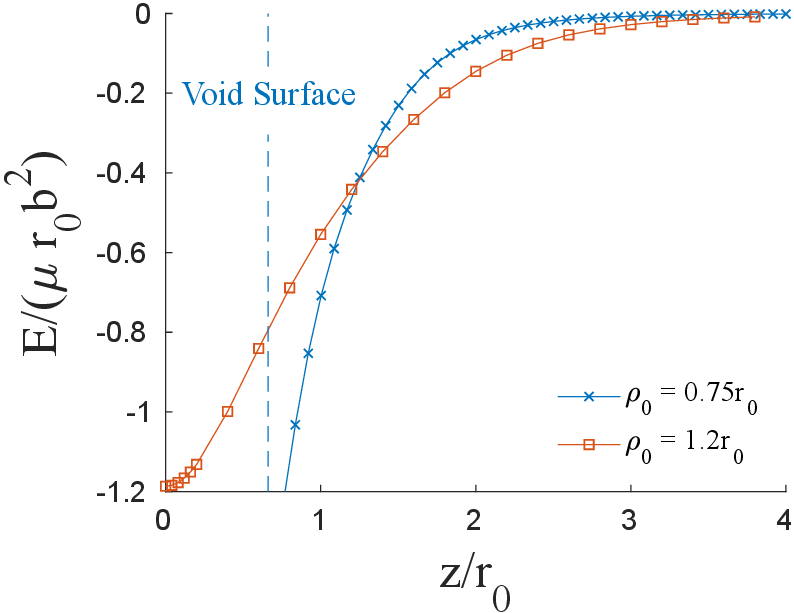}
	\caption{}
    \label{fig:result-void-pdl-b}
\end{subfigure}
\caption{The simulation result for PDLs whose radius $\rho_0$ are 0.75 and 1.2 times the void radius $r_0$. (a) Normalized glide forces $f_zr_0/(\mu b^2)$.  The inset shows sketches of the geometry, and the arrows indicates the positive direction of the force $f_z$. (b) Normalized interaction energy $E_{\rm el}^{\rm img}/(\mu r_0b^2)$ between the void and the dislocation loop. The dashed line indicates the void surface when $\rho_0=0.75r_0$.}
\label{fig:result-void-pdl}
\end{figure}

Figure~\ref{fig:result-void-pdl-a} shows the glide forces of the PDL with different radius values ($0.75\,r_0$ and $1.2\,r_0$).  The dislocations are always attracted towards the center of the void.  For the case of the PDL with radius less than the void, the dislocation loop accelerates when getting closer to the void surface. The force reaches negative infinity when the loop is approaching the void surface. 
For the case of the PDL with the radius larger than the void, the dislocation loop initially accelerates as it moves from far away towards the void surface. Then, the loop decelerates and finally stops at $z=0$, forming a ``Saturn ring'' around the void. The results are consistent with the qualitative analyses by \citeauthor{willis1969dislocation_loop}~(\citeyear{willis1969dislocation_loop}).

In addition, the image elastic energy is evaluated during the simulation, as shown in figure~\ref{fig:result-void-pdl-a}. 
The interaction energy for the loop with radius $\rho_0 = 0.75\,r_0$ keeps decreasing towards negative infinity as the dislocation loop approaches the void surface, while the interaction energy for the loop with radius $\rho_0 = 1.2\,r_0$ radius reaches a minimum at $z=0$. These results are consistent with the behaviors of the PDL in the simulation. We have also verified (by numerical differentiation) that the glide force and the image elastic energy satisfy the following relation:
\begin{equation}
f_z(z) \cdot 2\pi\rho_0\equiv F^{\rm img}_z(z)=-\frac{\partial E_{\rm el}^{\rm img}(z)}{\partial z}
\end{equation}
where $F^{\rm img}_z$ is the total glide force on the dislocation loop due to the image stress.

\section{Discussions and conclusions}
\label{sec:discussion}

In the present work, we develop an image stress solver for arbitrary tractions on spherical voids. Based on the expansion of the displacement potentials in the Papkovich-Neuber representation of the isotropic elasticity problem, the image stress solver outperforms the finite-element method (FEM) in terms of both efficiency and accuracy.
Our method is also expected to be more efficient than the boundary element method (BEM). The stiffness matrix in BEM is usually a dense matrix, which implies that the time to solve the linear system grows with the square of the matrix size (number of boundary element nodes). However, $F_J^K$ is a sparse matrix in our method, thus the time to solve the linear system grows only linearly with the matrix size (number of fundamental bases).

We have demonstrated that the spherical harmonics image stress solver can be coupled with the DD model to simulate void-dislocation interaction. The method can also be readily generalized for solving other elasticity problems with spherical boundaries. For example to model the growth of Helium bubbles inside irradiated metals (\citeauthor{misra2007helium_bubble}, \citeyear{misra2007helium_bubble}), the boundary condition can be changed to constant pressure (instead of zero traction) on the spherical surface. In addition, by changing the radius-dependent factor $1/r^{l+1}$ to $r^l$ when constructing the fundamental basis $\bm{\psi}^{(K)}$, the method can be used for solving the contact problem between spheres.  Keeping both $1/r^{l+1}$ and $r^l$ terms would enable efficient solutions for dislocation interaction with inclusions and inhomogeneities. 

\section*{Acknowledgements}

The authors are grateful for the help from Dr. Chao Pu (AK Steel Corp) in setting up the dislocation-void interaction FEM simulation in ABAQUS, and useful discussions about the spherical harmonics analysis with Dr. Yang Zhou (Department of Mathematics, Stanford University). This work was supported by the U.S. Department of Energy, Office of Basic Energy Sciences, Division of Materials Sciences and Engineering under Award No. DE-SC0010412.

\appendix
\section{Spherical harmonic coefficients for fundamental elasticity solutions}
\label{adx:detail_derivation}

In section~\ref{sec:method}, we express the displacement potential $\bm{\psi}^{(K)}$, the corresponding $\bm{u}^{(K)}$, $\bm{\sigma}^{(K)}$ and $\bm{T}^{(K)}$ fields in terms of the product of a pre-factor $f(r)$ and a radius-independent part $U^K_j(\theta,\varphi)$, $S^K_{ij}(\theta,\varphi)$, $F^K_j(\theta,\varphi)$ in equations~(\ref{eqn:displacement_solution},\ref{eqn:stress_solution},\ref{eqn:traction_solution}).  These radius independent functions can then be represented as linear combinations of spherical harmonic functions, for example equation~(\ref{eqn:SH_decompose_U}). In this section, we will show how we obtain the expansion coefficients $\hat{U}^K_{jl'm'}$, $\hat{S}^K_{ijl'm'}$, $\hat{F}^K_{jl'm'}$ starting from the spherical harmonics representation of displacement potential $\bm{\psi}^{(K)}$. These functions are implemented in ShElastic toolbox and the coefficients up to $l_{\max}=40$ are pre-calculated and saved as a sparse matrix.

\subsection{Completeness of spherical harmonics solution}
\label{adx:harmonic_potential}

The Papkovich-Neuber solution is a general solution to the equilibrium condition (\ref{eqn:eq_cond}) without rigid-body motion and rotation (\citeauthor{barber2009elasticity} 3rd ed., \citeyear{barber2009elasticity}), and can be written in terms of displacement potential vector $\bm{\psi}$ and function $\phi$:

\begin{equation}
2\mu\bm{u} = -4(1-\nu)\bm{\psi} + \nabla(\bm{r\cdot\psi}+\phi)
\label{eqn:Papkovich-Neuber}
\end{equation}

\noindent
where both $\bm{\psi}$ and $\phi$ are harmonic, i.e., satisfying Laplace's equation $\nabla^2\psi_j=0,(j=x,y,z);\,\nabla^2\phi=0$. Similar to section~\ref{sec:decompose_modes}, 
we can choose the basis functions for the displacement potentials $\bm{\psi}^{(k,l,m)}$ and $\phi^{(l',m')}$ as the general solution to the Laplace's equation:

\begin{eqnarray}
\bm{\psi}^{(k,l,m)}\equiv
\frac{1}{r^{l+1}}\hat{\bm{e}}_kY_l^m(\theta,\varphi), \quad
\phi^{(l',m')}\equiv
\frac{1}{r^{l'+1}}Y_{l'}^{m'}(\theta,\varphi)
\end{eqnarray}

\noindent
so that they completely span the solution space of the spherical void problem. However, in equation~(\ref{eqn:displacement_mode}), we only choose the $\bm{\psi}$ terms as the fundamental bases.
In the following, We will show  that such an expansion is still complete, by proving that $\phi^{(l',m')}$ can be written as a linear combination of $\bm{r}\cdot\bm{\psi}^{(k,l,m)}$.

Given the definition of $\bm{\psi}^{(k,l,m)}$ and $\phi^{(l',m')}$, we have
\begin{equation}
\bm{r}\cdot\bm{\psi}^{(k,l,m)} = x_{k} \frac{Y_l^m} { r^{l+1} } \, , \quad
\phi^{(l-1,m)} = \frac{Y_{l-1}^m}{r^{l}}\quad (l\ge 1)
\end{equation}
We will show that $\phi^{(l-1,m)}$ can be written as a linear combination of $\bm{r}\cdot\bm{\psi}^{(k,l,m)}$.
Because $Y_1^1 = \sqrt{3/2}\,(x+iy)/r$ and $Y_1^0 = \sqrt{3}\,z/r$ (see equations(\ref{eqn:Y1-1_expression})-(\ref{eqn:Y11_expression})),
we have (for $-l+1 \le m \le l$)
\begin{equation}
\begin{split}
\begin{bmatrix}
  \bm{r\cdot\psi}^{(x,l,m-1)} + i\bm{r\cdot\psi}^{(y,l,m-1)}\\
  \sqrt{2}\, \bm{r\cdot\psi}^{(z,l,m)}
\end{bmatrix}
&=
\frac{1}{r^l}\sqrt{\frac{2}{3}}
\begin{bmatrix}
  Y_l^{m-1}Y_1^{1} \\
  Y_l^mY_1^0
\end{bmatrix}
\end{split}
\label{eqn:psi_to_product}
\end{equation}
In \ref{adx:spherical_harmonics} we show how the products of spherical harmonic functions can be written as linear combinations of $Y_l^m$.  
Specifically, equation (\ref{eqn:linear_combination_vsh1}) leads to the following linear equations:
\begin{equation}
\begin{split}
\begin{bmatrix}
  Y_l^{m-1}Y_1^{1} \\
  Y_l^mY_1^0
\end{bmatrix}
=\bm{C^-}
\begin{bmatrix}
  Y_{l+1}^{m} \\ Y_{l-1}^{m}
\end{bmatrix}
&\equiv
\begin{bmatrix}
  C_{(l,1)}^{(m-1,1)} & C_{(l,-1)}^{(m-1,1)} \\
  C_{(l,1)}^{(m,0)} & C_{(l,-1)}^{(m,0)}
\end{bmatrix}
\begin{bmatrix}
  Y_{l+1}^{m} \\ Y_{l-1}^{m}
\end{bmatrix}
\end{split}
\label{eqn:linear_equation_phi_-}
\end{equation}
Because of the orthogonality of Clebsch-Gordan coefficients (\citeauthor{clebsch-gordan-coefficients}, \citeyear{clebsch-gordan-coefficients}), $\bm{C^-}$ is a nonsingular matrix, i.e. it can be inverted. Therefore (for $-l+1 \le m \le l$), we can obtain $Y_{l-1}^m$ by solving the equation (\ref{eqn:linear_equation_phi_-}):
\begin{equation}
\phi^{(l-1,m)}=\frac{Y_{l-1}^m}{r^l}=\frac{1}{\left\vert\bm{C}^-\right\vert r^l}
	\left[C_{(l,-1)}^{(m,0)}(Y_l^m Y_1^0)-C_{(l,1)}^{(m,0)}(Y_l^{m-1}Y_1^{1})\right]
\end{equation}
where $\left\vert\bm{C^-}\right\vert$ is the determinant of matrix $\bm{C^-}$.
Given equation (\ref{eqn:psi_to_product}), $\phi^{(l-1,m)}$ is now expressed as a linear combination of 
$\bm{r\cdot\psi}^{(x,l,m-1)}$, $\bm{r\cdot\psi}^{(y,l,m-1)}$, and $\bm{r\cdot\psi}^{(z,l,m)}$.

Similarly, for $-l \le m \le l-1$ we have
\begin{equation}
\begin{split}
\begin{bmatrix}
  \bm{r\cdot\psi}^{(x,l,m+1)} - i\bm{r\cdot\psi}^{(y,l,m+1)}\\
  \sqrt{2}\, \bm{r\cdot\psi}^{(z,l,m)}
\end{bmatrix}
&=
\frac{1}{r^l}\sqrt{\frac{2}{3}}
\begin{bmatrix}
  Y_l^{m+1}Y_1^{-1} \\
  Y_l^mY_1^0
\end{bmatrix}
\end{split}
\end{equation}
where
\begin{equation}
\begin{split}
\begin{bmatrix}
  Y_l^{m+1}Y_1^{-1} \\
  Y_l^mY_1^0
\end{bmatrix}
=\bm{C^+}
\begin{bmatrix}
  Y_{l+1}^{m} \\ Y_{l-1}^{m}
\end{bmatrix}
&\equiv
\begin{bmatrix}
  C_{(l,1)}^{(m+1,-1)} & C_{(l,-1)}^{(m+1,-1)} \\
  C_{(l,1)}^{(m,0)} & C_{(l,-1)}^{(m,0)}
\end{bmatrix}
\begin{bmatrix}
  Y_{l+1}^{m} \\ Y_{l-1}^{m}
\end{bmatrix}
\end{split}
\label{eqn:linear_equation_phi_+}
\end{equation}
which allows us to show that
\begin{equation}
\phi^{(l-1,m)}=\frac{Y_{l-1}^m}{r^l}=\frac{1}{\left\vert\bm{C^+}\right\vert r^l}
	\left[C_{(l,-1)}^{(m,0)}(Y_l^m Y_1^0)-C_{(l,1)}^{(m,0)}(Y_l^{m+1}Y_1^{-1})\right]
\end{equation}
so that $\phi^{(l-1,m)}$ is now expressed as a linear combination of 
$\bm{r\cdot\psi}^{(x,l,m+1)}$, $\bm{r\cdot\psi}^{(y,l,m+1)}$, and $\bm{r\cdot\psi}^{(z,l,m)}$.
This proves that the expansion of the vector potential $\bm{\psi}$ alone into spherical harmonics completely spans the solution space.

\subsection{Displacement solution}
\label{adx:derive_displacement}

In this section, we will show how we obtain the radius independent part of the displacement basis $U^K_j(\theta,\varphi)$ and thus the expansion coefficients $\hat{U}^K_{jl'm'}$. Based on the fundamental bases of the displacement potential function (\ref{eqn:displacement_potential_mode}), the displacement solution $\bm{u}^{(k,l,m)}=\bm{u}^{(K)}$ can be written as:
\begin{equation}
2\mu\bm{u}^{(K)}= 
-4(1-\nu)\hat{\bm{e}}_{k}\frac{Y_l^m}{r^{l+1}}+
  \nabla\left(\frac{x_k \, {Y}_l^m }{r^{l+1}}\right)
\end{equation}

The first term is already represented in terms of spherical harmonic functions. To expand the second term, it is convenient to introduce the first and second vector spherical harmonics (VSH), $\bm{Y}_{l}^{m}$ and $\bm{\Psi}_{l}^{m}$,

\begin{align}
\bm{Y}_{l}^{m} &\equiv Y_l^m \hat{\bm{r}} \label{eqn:definition_VSH1}\\
\bm{\Psi}_{l}^{m} &\equiv r\bm{\nabla}Y_l^m \label{eqn:definition_VSH2}
\end{align}
where $\hat{\bm{r}}=[x/r,\,y/r,\,z/r]^T$ is the unit vector in the radial direction. 
Both $\bm{Y}_{l}^{m}$ and $\bm{\Psi}_{l}^{m}$ are independent of $r$, so they can be 
expressed as linear combinations of $Y_l^m$, as shown in equations (\ref{eqn:first_vsh_final}) and (\ref{eqn:second_vsh_final}).

For brevity, we introduce coefficients $Q_{klm}^{l'm'}$ to describe the relationship between $\bm{Y}_{l}^{m}$ and $Y_{l}^{m}$,
\begin{equation}
 [\bm{Y}_l^m(\theta,\varphi)]_k = \sum_{l',m'} Q_{klm}^{l'm'} \, Y_{l'}^{m'}(\theta,\varphi)
\end{equation}
where $[\bm{Y}_l^m(\theta,\varphi)]_k$ is the $k$-th component of $\bm{Y}_l^m$.
Given these definitions, we can rewrite the displacement solution as
\begin{align}
2\mu\,u_j^{(K)} &= 
-4(1-\nu)\,\delta_{jk}\frac{1}{r^{l+1}}\,Y_l^m + \frac{\partial}{\partial x_j}\left(\frac{1}{r^l} \, [\bm{Y}_l^m]_k \right) \nonumber \\
  &= 
-4(1-\nu)\,\delta_{jk}\frac{1}{r^{l+1}}\,Y_l^m + \sum_{l',m'} Q_{klm}^{l'm'} \, \frac{\partial}{\partial x_j}\left(\frac{1}{r^l} \, Y_{l'}^{m'} \right)  \\
  &=
-4(1-\nu)\,\delta_{jk}\frac{1}{r^{l+1}}\,Y_l^m + \sum_{l',m'} Q_{klm}^{l'm'} \, \left(\frac{-l}{r^{l+1}} \, [\bm{Y}_{l'}^{m'}]_j + \frac{1}{r^{l+1}} [\bm{\Psi}_{l'}^{m'}]_j \right) 
\nonumber
\end{align}
Therefore, the $r$-independent part of the displacement solution is
\begin{align}
U_j^{K} &\equiv 2\mu\,r^{l+1}\,u_j^{(K)} 
 = \sum_{l',m'} \hat{U}_{jl'm'}^K \, Y_{l'}^{m'}
\nonumber \\ 
&= -4(1-\nu)\,\delta_{jk}\,Y_l^m + \sum_{l',m'} Q_{klm}^{l'm'} \, \left( -l\,[\bm{Y}_{l'}^{m'}]_j + [\bm{\Psi}_{l'}^{m'}]_j \right) 
\end{align}
Given that the VSH functions $\bm{Y}_l^m$ and $\bm{\Psi}_l^m$ can be written as linear combination of $Y_l^m$ according to equations (\ref{eqn:first_vsh_final}) and (\ref{eqn:second_vsh_final}), we now have expressed $U_j^{K}$ as linear combination of spherical harmonic functions.

\subsection{Stress and traction solution}

The spatial gradient of the displacement solution $u^{(K)}_{j}$ can be expressed as
\begin{equation}
\begin{split}
u^{(K)}_{j,i} &= 
\sum_{l',m'}\hat{U}^{klm}_{jl'm'}
    \frac{\partial}{\partial x_i} \left( \frac{Y_{l'}^{m'}}{2\mu \, r^{l+1}} \right)\\
    &=
\frac{1}{2\mu r^{l+2}} \sum_{l',m'} \hat{U}^{klm}_{jl'm'}
    \left[-(l+1)[\bm{Y}_{l'}^{m'}]_i
          +[\bm{\Psi}_{l'}^{m'}]_i\right]\\
\end{split}
\label{eqn:evaluate_displacement}
\end{equation}
Similarly, given that the VSH $\bm{Y}_l^m$ and $\bm{\Psi}_l^m$ can be written as linear combination of $Y_l^m$ according to equations (\ref{eqn:first_vsh_final}) and (\ref{eqn:second_vsh_final}), we now have expressed the displacement gradient $u_{j,i}^{(K)}$ as linear combination of $Y_l^m$.
The stress field can be calculated through the following constitutive relation: 

\begin{equation}
\sigma_{ij}^{(K)} = 
\lambda \, u^{(K)}_{u,u}\delta_{ij} + \mu\, (u^{(K)}_{i,j}+u^{(K)}_{j,i})
\label{eqn:constitutive_decomposition}
\end{equation}

\noindent
This allows us to express the $r$-independent part of the stress field as a linear combination of $Y_l^m$,
\begin{equation}
S^{(K)}_{ij} \equiv r^{l+2} \, \sigma_{ij}^{(K)} =
  \sum_{l',m'}\hat{S}^{klm}_{ijl'm'}Y_{l'}^{m'}
\end{equation}
Finally, the traction vector of mode $(K)$ at the void surface ($r=r_0$) can be evaluated as:
\begin{equation}
T_j^{(K)}=\sigma_{ij}^{(K)}n_i=
-\frac{x_i}{r}\sigma_{ij}^{(K)}=
-\frac{1}{r_0^{l+2}}\sum_{l',m'}\sum_i
\hat{S}^{klm}_{ijl'm'}[\bm{Y}_{l'}^{m'}]_i
\end{equation}
where $\bm{n}=-\hat{\bm{r}}$ is the normal vector of the void surface.
Therefore,
\begin{align}
F_j^{(K)} &\equiv r_0^{l+2}\, T_j^{(K)} = 
	\sum_{l',m'} \hat{F}_{jl'm'}^{klm} \, Y_{l'}^{m'} \nonumber \\
 &= -\sum_{l',m'}\sum_i \hat{S}^{klm}_{ijl'm'} \, [\bm{Y}_{l'}^{m'}]_i 
  = -\sum_{l',m'}\sum_{l'',m''}\sum_i \hat{S}^{klm}_{ijl'm'} \, Q_{il'm'}^{l''m''} \, Y_{l''}^{m''}
\end{align}

\section{Transformation rules for spherical harmonic functions}
\label{adx:spherical_harmonics}

In \ref{adx:detail_derivation}, we saw that obtaining the coefficients $\hat{U}^K_J$, $\hat{S}^K_I$, $\hat{F}^K_J$ requires the expansion of VSH functions $\bm{Y}_l^m$, $\bm{\Psi}_l^m$ in terms of $Y_l^m$.  In this appendix, we provide explicit expressions for these expansion coefficients and the proof of the expansion. We start with the explicit expressions of the spherical harmonic functions used in this paper, since there are multiple conventions of $Y_l^m$ in the literature.

\subsection{Definitions of spherical harmonic functions}
\label{adx:convention}

$Y_l^m(\theta,\varphi)$ are the complex spherical harmonic functions defined on the spherical surface, where $\theta \in [0,\pi]$ is the colatitude angle and the $\varphi \in [0, 2\pi]$ is the longitude angle, $l = 0, 1, \cdots$, and $m = -l, -l+1, \cdots, l$.

\begin{equation}
Y_l^m(\theta,\varphi)=
\begin{dcases}
\sqrt{(2l+1)\frac{(l-m)!}{(l+m)!}}P_{lm}(\cos\theta)e^{im\varphi},& m\geq0\\
(-1)^{|m|} \left[Y_l^{|m|}(\theta,\varphi)\right]^{*}, & m<0
\end{dcases}
\end{equation}

\noindent
where $\left[Y_l^{-|m|}\right]^*$ is the complex conjugate function of $Y_l^{-|m|}$, and $P_{lm}(\mu)$ is the associated Legendre function (without Condon-Shortley phase), which is defined as:

\begin{eqnarray}
P_{lm}(\mu)=(1-\mu^2)^{m/2}\frac{\text{d}^m}{\text{d}\mu^m}P_l(\mu) \\
P_l(\mu)=\frac{1}{2^ll!}\frac{\text{d}^l}{\text{d}\mu^l}(\mu^2-1)^l
\end{eqnarray}

Some examples of the low-order complex spherical harmonics are listed here in both spherical and Cartesian coordinates:

\begin{align}
Y_0^0 &= & & 1 &&& \\
Y_1^{-1} &= &-& \sqrt{\frac{3}{2}}e^{-i\varphi}\sin\theta & &= 
            &-&\sqrt{\frac{3}{2}}\frac{x-iy}{r} 
	\label{eqn:Y1-1_expression}\\
Y_1^0 &= & & \sqrt{3}\cos\theta & &= & & \sqrt{3}\frac{z}{r} 
	\label{eqn:Y10_expression}\\
Y_1^1 &= & & \sqrt{\frac{3}{2}}e^{i\varphi}\sin\theta
    & &= & & \sqrt{\frac{3}{2}}\frac{x+iy}{r}   
    \label{eqn:Y11_expression}\\
Y_2^{-2} &= & & \frac{\sqrt{30}}{4}e^{-2i\varphi}\sin^2\theta 
       & &= & & \frac{\sqrt{30}}{4}\frac{(x-iy)^2}{r^2} \\
Y_2^{-1} &= &-& \sqrt{\frac{15}{2}}e^{-i\varphi}\sin\theta\cos\theta 
	   & &= &-&\sqrt{\frac{15}{2}}\frac{(x-iy)z}{r^2} \\
Y_2^0 &= & & \frac{\sqrt{5}}{2}(3\cos^2\theta-1) 
	& &= & & \frac{\sqrt{5}}{2}\frac{2z^2-x^2-y^2}{r^2} \\
Y_2^1 &= & & \sqrt{\frac{15}{2}}e^{i\varphi}\sin\theta\cos\theta 
	& &= & & \sqrt{\frac{15}{2}}\frac{(x+iy)z}{r^2} \\
Y_2^2 &= & & \frac{\sqrt{30}}{4}e^{2i\varphi}\sin^2\theta 
	& &= & & \frac{\sqrt{30}}{4}\frac{(x+iy)^2}{r^2}
\end{align}

\noindent
where the radius $r=\sqrt{x^2+y^2+z^2}$.

Two spherical harmonics $Y_l^{m}$ and $Y_{l'}^{m'}$ satisfy the orthogonality condition with  $4\pi$-Geodesy normalization:

\begin{equation}
\int_\Omega [Y_l^{m}(\theta,\varphi)]^*Y_{l'}^{m'}(\theta,\varphi)\text{d}\Omega = 4\pi \delta_{ll'} \delta_{mm'}
\label{eqn:orthogonality_condition}
\end{equation}

The product of two spherical harmonic functions $Y_{l_1}^{m_1}$, $Y_{l_2}^{m_2}$ can be expanded as a linear combination of spherical harmonics:

\begin{eqnarray} \nonumber
Y_{l_1}^{m_1}(\theta,\varphi)Y_{l_2}^{m_2}(\theta,\varphi) &=& 
    \sum_{L=\vert l_1-l_2\vert}^{\vert l_1+l_2\vert}
    c(l_1,m_1,l_2,m_2,L,M)c(l_1,0,l_2,0,L,0) \\
  &&  \sqrt{\frac{(2l_1+1)(2l_2+1)}{2L+1}}\, Y_L^M(\theta,\varphi)
\label{eqn:sh_product}
\end{eqnarray}

\noindent
where $M=m_1+m_2$, and $c(l_1,m_1,l_2,m_2,L,M)$ is the Clebsch-Gordan (CG) coefficients (\citeauthor{clebsch-gordan-coefficients}, \citeyear{clebsch-gordan-coefficients}). Note that the CG coefficients for invalid indices such as $M > L$ or $M < -L$ are zero. 

\subsection{Expansion into spherical harmonic functions}
\label{adx:expansion}

Any complex square-integrable function on the unit sphere can be expressed in terms of the linear combination of spherical harmonics:

\begin{equation}
\Omega(\theta,\varphi)=\sum_{l=0}^{\infty}\sum_{m=-l}^{+l}\hat{\Omega}_{lm}Y_l^m(\theta,\varphi)
\label{eqn:sh_expand}
\end{equation}

\noindent
where $\hat{\Omega}_{lm}$ are the complex spherical harmonic coefficients.

For any complex function in 3D space $\omega(r, \theta,\varphi)$, the spherical harmonic expansion can be written as:

\begin{equation}
\omega(r,\theta,\varphi)=\sum_{l=0}^{\infty}\sum_{m=-l}^{+l}\omega_{lm}(r)Y_l^m(\theta,\varphi)
\label{eqn:sh_expand_3d}
\end{equation}

\noindent
where the coefficients $\omega_{lm}(r)$ are functions of $r$ instead of constants. 

SHTOOLS (\citeauthor{shtools}, \citeyear{shtools}) provides fast forward and backward spherical harmonics transform in the form of equation~(\ref{eqn:sh_expand}). The forward transformation is expanding a discretized 2D-surface function $\Omega(\theta,\varphi)$ to spherical harmonic coefficients $\hat{\Omega}_{lm}$, while the backward transformation is reconstructing the 2d-surface function from the coefficients. For 3D-functions, the spherical harmonic transform can be performed at a specific radius $r=r_0$:

\begin{equation}
\omega(r_0,\theta,\varphi)=\sum_{l=0}^{\infty}\sum_{m=-l}^{+l}\hat{\omega}_{lm}(r_0)Y_l^m(\theta,\varphi)
\label{eqn:sh_expand_3d_r0}
\end{equation}

\subsection{Vector spherical harmonics}
\label{adx:vsh}

In this work, we make use of vector spherical harmonic functions to simplify the notation. We follow the convention in \citeauthor{VSH} (\citeyear{VSH}): given a scalar spherical harmonic function $Y_l^m(\theta,\varphi)$, the first and second vector spherical harmonics (VSH) are defined by equations (\ref{eqn:definition_VSH1}) and (\ref{eqn:definition_VSH2}), respectively.
Note that both $\bm{Y}_l^m$ and $\bm{\Psi}_l^m$ are functions of $\theta$ and $\varphi$ only, i.e. they are independent of $r$, similar to $Y_l^m$.

In this section, we show that the first and second VSH functions, $\bm{Y}_l^m$ and $\bm{\Psi}_l^m$, can be expressed as linear combinations of $Y_l^m$ as follows:

\begin{equation}
\bm{Y}_l^m=
\begin{bmatrix}
\sqrt\frac{1}{6}\, (C_{(l,1)}^{(m, 1)}Y_{l+1}^{m+1} + C_{(l,-1)}^{(m, 1)}Y_{l-1}^{m+1})
                  - C_{(l,1)}^{(m,-1)}Y_{l+1}^{m-1} - C_{(l,-1)}^{(m,-1)}Y_{l-1}^{m-1}) \\
\sqrt\frac{1}{6}\,i(C_{(l,1)}^{(m, 1)}Y_{l+1}^{m+1} + C_{(l,-1)}^{(m, 1)}Y_{l-1}^{m+1}
				  + C_{(l,1)}^{(m,-1)}Y_{l+1}^{m-1} + C_{(l,-1)}^{(m,-1)}Y_{l-1}^{m-1}) \\
\sqrt\frac{1}{3}\, (C_{(l,1)}^{(m, 0)}Y_{l+1}^{m}   + C_{(l,-1)}^{(m, 0)}Y_{l-1}^{m})
\end{bmatrix}
\label{eqn:first_vsh_final}
\end{equation}

\begin{equation}
\bm{\Psi}_l^m(\theta,\varphi)=(l+1)\bm{Y}_l^m + \frac{1}{2a_l^m}
\begin{bmatrix}
a_{l+1}^{m-1}Y_{l+1}^{m-1} - a_{l+1}^{m+1}Y_{l+1}^{m+1} \\
i(a_{l+1}^{m-1}Y_{l+1}^{m-1} + a_{l+1}^{m+1}Y_{l+1}^{m+1}) \\
-2a_{l+1}^{m}Y_{l+1}^m
\end{bmatrix}
\label{eqn:second_vsh_final}
\end{equation}
where
\begin{equation}
a_l^m \equiv \sqrt{\frac{(l+m)!\,(l-m)!}{2l+1}}
\label{eqn:alm}
\end{equation} 

\subsubsection{Proof of the first vector spherical harmonics expression}
Based on equations (\ref{eqn:Y1-1_expression})-(\ref{eqn:Y11_expression}), the unit vector $\hat{\bm{r}}$ can be written in terms of $Y_l^m$ as:

\begin{equation}
\hat{\bm{r}} = 
\begin{bmatrix}
	x/r\\y/r\\z/r
\end{bmatrix}=\sqrt{\frac{1}{6}}
\begin{bmatrix}
Y_1^1-Y_1^{-1} \\ i(Y_1^{1}+Y_1^{-1}) \\ \sqrt{2}Y_1^0
\end{bmatrix}
\end{equation}
Therefore, the first VSH function $\bm{Y}_l^m(\theta,\phi)$ can be written as:
\begin{equation}
\bm{Y}_l^m(\theta,\phi)\equiv Y_l^m\hat{\bm{r}}=
\sqrt\frac{1}{6}
\begin{bmatrix}
Y_1^1Y_l^m-Y_1^{-1}Y_l^m \\
i(Y_1^{1}Y_l^m+Y_1^{-1}Y_l^m) \\
\sqrt{2}Y_1^0Y_l^m
\end{bmatrix}
\label{eqn:first_vsh}
\end{equation}

\noindent
where the product of spherical harmonic functions can be further expressed as linear combination of spherical harmonic functions using equation~(\ref{eqn:sh_product}). Specifically,

\begin{equation}
\begin{split}
Y_l^mY_1^1&=C_{(l,1)}^{(m,1)}Y_{l+1}^{m+1}+C_{(l,0)}^{(m,1)}Y_{l}^{m+1}+C_{(l,-1)}^{(m,1)}Y_{l-1}^{m+1} \\
Y_l^mY_1^{-1}&=C_{(l,1)}^{(m,-1)}Y_{l+1}^{m-1}+C_{(l,0)}^{(m,-1)}Y_{l}^{m-1}+C_{(l,-1)}^{(m,-1)}Y_{l-1}^{m-1} \\
Y_l^mY_1^0&=C_{(l,1)}^{(m,0)}Y_{l+1}^{m}+C_{(l,0)}^{(m,0)}Y_{l}^{m}+C_{(l,-1)}^{(m,0)}Y_{l-1}^{m}
\end{split}
\label{eqn:linear_combination_vsh1}
\end{equation}

\noindent where the constants $C_{(l,l')}^{(m,m')},\;(l',m'=0,\pm1)$ are:

\begin{equation}
\begin{split}
C_{(l,1)}^{(m,\pm1)}&=\sqrt{\frac{3(l+1)}{2l+3}}c(l,m,1,\pm1,l+1,m\pm1) \\
C_{(l,1)}^{(m,0)}&=\sqrt{\frac{3(l+1)}{2l+3}}c(l,m,1,0,l+1,m) \\
C_{(l,-1)}^{(m,\pm1)}&=\sqrt{\frac{3l}{2l-1}}c(l,m,1,\pm1,l-1,m\pm1) \\
C_{(l,-1)}^{(m,0)}&=\sqrt{\frac{3l}{2l-1}}c(l,m,1,0,l-1,m) \\
C_{(l,0)}^{(m,0)}&=C_{(l,0)}^{(m,\pm1)}=0
\end{split}
\label{eqn:coefficients_vsh1}
\end{equation}

\noindent
which leads to equation (\ref{eqn:first_vsh_final}) that we set out to prove.

\subsubsection{Proof of the second vector spherical harmonics expression}

In order to prove equation (\ref{eqn:second_vsh_final}), we introduce the irregular solid harmonic function $\mathcal{I}_l^m(r,\theta,\varphi)$, which is defined as:
\begin{equation}
\mathcal{I}_l^m(r,\theta,\varphi) \equiv 
\begin{dcases}
\frac{(l-m)!}{r^{l+1}}P_{lm}(\cos\theta)e^{im\varphi} \, , & m \ge 0
\\
(-1)^{|m|}\,\left[\mathcal{I}_l^{|m|}(r,\theta,\varphi)\right]^* \, , & m<0
\end{dcases}
\label{eqn:defIlm}
\end{equation}
and $\mathcal{I}_l^m$ and $Y_l^m$ are related with:
\begin{equation}
\mathcal{I}_l^m = a_l^m\,\frac{Y_l^m}{r^{l+1}}
\label{eqn:coefficients_vsh2}
\end{equation}
where $a_l^m$ is defined in equation~(\ref{eqn:alm}).

The second VSH function $\bm{\Psi}_l^m$ can now be expressed in terms of $\bm{Y}_l^m$ and $\mathcal{I}_l^{m}$ as follows.
\begin{align}
\bm{\Psi}_l^m &\equiv r \nabla Y_l^m = \frac{r}{a_l^m} \nabla \left( r^{l+1} \, \mathcal{I}_l^m \right) 
 \nonumber\\
 &= \frac{r}{a_l^m} \left[ (l+1) \, r^l \, \hat{\bm{r}} \, \mathcal{I}_l^m  
   + r^{l+1} \nabla \mathcal{I}_l^m  \right]  \nonumber\\
 &= (l+1) \, \bm{Y}_l^m + \frac{r^{l+2}}{a_l^m} \, \nabla\mathcal{I}_l^m
 \label{eqn:deriv_Psilm}
\end{align}
The gradient of $\mathcal{I}_l^m$ can be obtained by Hobson's differentiation (\citeauthor{hobson1931theory}, \citeyear{hobson1931theory}):
\begin{equation}
D_1\mathcal{I}_l^m = \mathcal{I}_{l+1}^{m-1},\quad
D_2\mathcal{I}_l^m =-\mathcal{I}_{l+1}^{m+1},\quad
D_3\mathcal{I}_l^m =-\mathcal{I}_{l+1}^{m};
\end{equation}
where the differentiation operator $D_i$ is defined as:
\begin{equation}
D_1 \equiv \left(\frac{\partial}{\partial x}-i\frac{\partial}{\partial y}\right),\quad
D_2 \equiv \left(\frac{\partial}{\partial x}+i\frac{\partial}{\partial y}\right),\quad
D_3 \equiv \frac{\partial}{\partial z}.
\label{eqn:D123}
\end{equation}
which leads to the gradient of $\mathcal{I}_l^m$:
\begin{equation}
\nabla\mathcal{I}_l^m = \frac{1}{2}
\begin{bmatrix}
\mathcal{I}_{l+1}^{m-1} - \mathcal{I}_{l+1}^{m+1} \\
i(\mathcal{I}_{l+1}^{m-1} + \mathcal{I}_{l+1}^{m+1}) \\
-2\mathcal{I}_{l+1}^m
\end{bmatrix}
\label{eqn:gradIlm}
\end{equation}
Plugging equations (\ref{eqn:gradIlm}) and (\ref{eqn:defIlm}) into equation (\ref{eqn:deriv_Psilm}) leads to the proof of equation (\ref{eqn:second_vsh_final}). 

\section{Image elastic energy}
\label{adx:image_energy}

Here we derive the expression for the image elastic energy in equation~(\ref{eqn:image_elastic_energy}). 
For clarity, let us define $E_{\rm el}^\infty$ as the elastic energy of an infinite medium containing dislocations, as shown in figure~\ref{fig:decompose_problem}(a).  For simplicity, we shall assume that there are no dislocations in the spherical region corresponding to the hole.
Let $E_{\rm el}^{\rm tot}$ be the elastic energy of interest, for which a spherical hole is carved out of the elastic medium containing the dislocations, as shown in figure~\ref{fig:decompose_problem}(c).
The image elastic energy is the difference between these two energies, i.e.
\begin{equation}
  E_{\rm el}^{\rm img} = E_{\rm el}^{\rm tot} - E_{\rm el}^{\infty}
\end{equation}
The image elastic energy can be considered as the interaction energy between the dislocations and the spherical void.  We note that $E_{\rm el}^{\rm img}$ has a negative value, corresponding to the attractive nature of the dislocation-void interaction.

We emphasize that $E_{\rm el}^{\rm img}$ is \emph{not} the elastic energy $E_{\rm el}^{\rm (b)}$ of the image stress problem, which is shown in figure~\ref{fig:decompose_problem}(b).
In other words, while the stress field of Problem (c) in figure~\ref{fig:decompose_problem} can be obtained by superposing the stress fields of Problem (a) and Problem (b), the same is not true for elastic energies.
By the way, the elastic energy of Problem (b) has a positive value and can be expressed as
\begin{equation}
  E_{\rm el}^{\rm (b)} = \frac{1}{2} \int_{\partial\Omega} 
\bm{T}^{\rm img}\cdot \bm{u}^{\rm img}\;dS 
  = -\frac{1}{2} \int_{\partial\Omega} 
\bm{T}^{\infty}\cdot \bm{u}^{\rm img}\;dS 
\end{equation}

\begin{figure}[!ht]
\centering
\includegraphics[width=\linewidth]{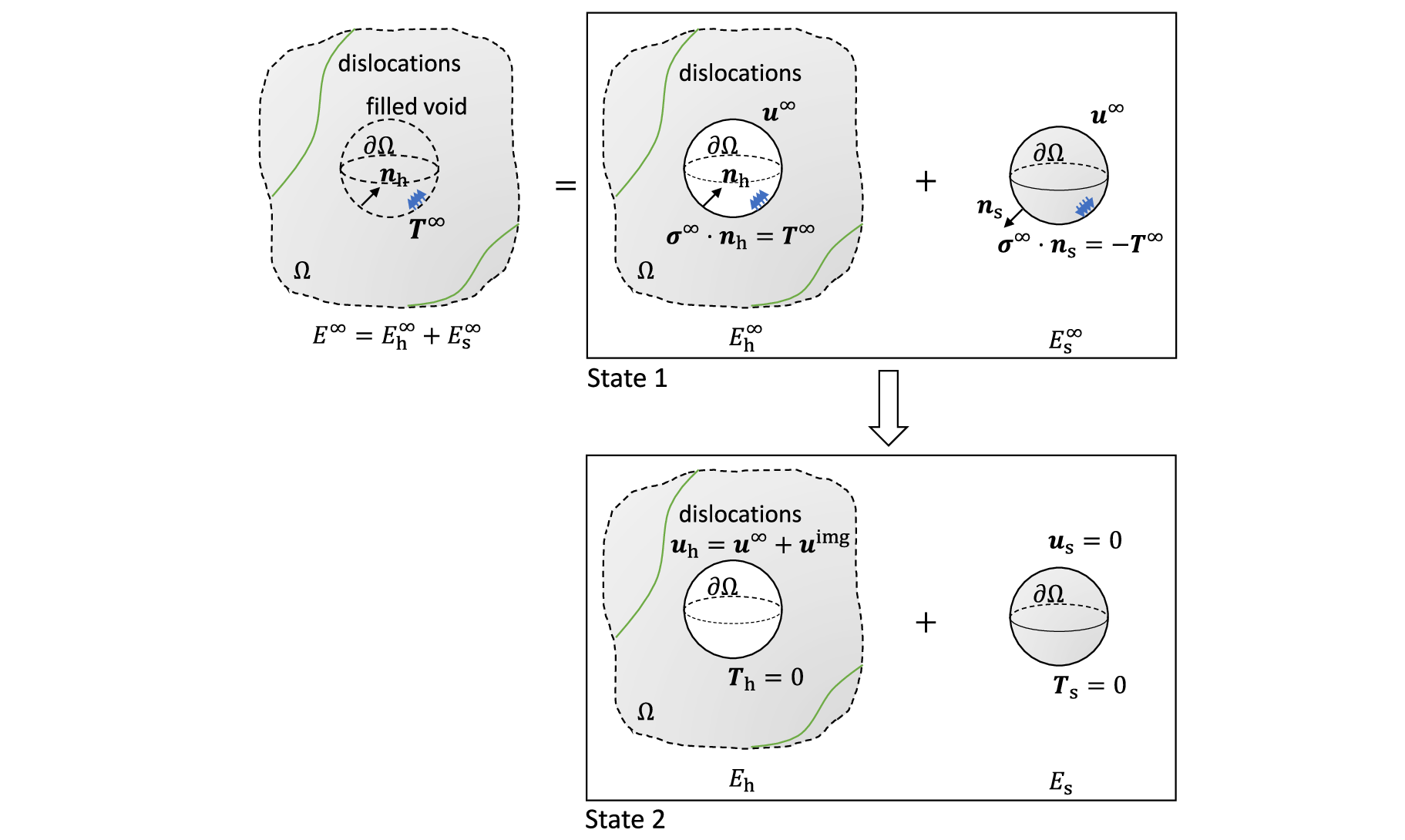}
\caption{A reversible path between two states.  In State 1, a spherical region is carved out of an infinite medium containing dislocations, but traction forces are applied to the surfaces of both the void and the sphere to maintain the stress field of dislocations in an infinite medium.
In State 2, the traction forces on both surfaces have been removed.}
\label{fig:reversible_path}
\end{figure}

To derive $E_{\rm el}^{\rm img}$, we consider a reversible path between two states shown in figure~\ref{fig:reversible_path}, and $E_{\rm el}^{\rm img}$ is the work done to the elastic medium along this path.
Starting from an infinite medium containing dislocations but no void, State 1 is obtained by cutting the spherical region out of the medium but maintaining tractions on the surfaces of both the void and the solid sphere.
The traction on the void surface is $\bm{T}^{\infty} = \bm{\sigma}^\infty\cdot\bm{n}_{\rm h}$, where $\bm{n}_{\rm h}$ is the surface normal of the void (pointing towards the spherical center); the traction on the solid sphere surface is $-\bm{T}^{\infty} = \bm{\sigma}^\infty\cdot\bm{n}_{\rm s}$, where $\bm{n}_{\rm s}$ is the surface normal of the sphere (pointing away from the spherical center).
The elastic energy of State 1 is the same as $E_{\rm el}^{\infty}$.

State 2 is obtained from State 1 by gradually reducing the magnitude of the tractions to zero, on both the hole and sphere surfaces.
On the hole surface, since the traction goes from $\bm{T}^{\infty}$ to $0$ as the displacement increments by $\bm{u}^{\rm img}$, the work done is,
\begin{equation}
 \Delta W_{\rm h} = \frac{1}{2} \int_{\partial \Omega} \bm{T}^{\infty}\cdot\bm{u}^{\rm img}\;dS
\end{equation}
On the solid sphere surface, since the traction goes from $-\bm{T}^{\infty}$ to $0$ as the displacement changes from $\bm{u}^{\infty}$ to $0$ (assuming no dislocations in the sphere), the work done is,
\begin{equation}
 \Delta W_{\rm s} = \frac{1}{2} \int_{\partial \Omega} \left(-\bm{T}^{\infty}\right)
      \cdot \left(-\bm{u}^{\infty}\right) \;dS
      = \frac{1}{2} \int_{\partial \Omega} \bm{T}^{\infty}\cdot\bm{u}^{\infty} \;dS
\end{equation}
Therefore,
\begin{equation}
 E_{\rm el}^{\rm img} = \Delta W_{\rm h} + \Delta W_{\rm s} 
      = \frac{1}{2} \int_{\partial \Omega} \bm{T}^{\infty}
      \cdot \left(\bm{u}^\infty + \bm{u}^{\rm img}\right) \;dS
\end{equation}
which proves equation~(\ref{eqn:image_elastic_energy}).
The negative derivative of the image elastic energy $E_{\rm el}^{\rm img}$ with respect to the dislocation position gives the attractive force exerted by the void on the dislocation, which (for self-consistency) is identical to the Peach-Koehler force of the image stress, as demonstrated in Section~\ref{sec:case_void_PDL_interaction}.

\section{Boundary value problems}
\label{adx:BVP}

{In this section, we provide the definitions of the exact boundary value problems in} figure~\ref{fig:decompose_problem}. {For isotropic homogeneous materials in the absence of body forces, the governing equation for the problem shown in} figure~\ref{fig:decompose_problem}(c) is:
\begin{equation} \begin{aligned}
    \mu \, u_{j,kk}+(\lambda+\mu)\, u_{k,kj}=0,
\end{aligned} \end{equation}
where $u_j$ is the displacement field, from which the stress field can be obtained as,
\begin{equation} \begin{aligned}
\sigma_{ij}=\lambda\delta_{ij}u_{k,k}+\mu(u_{i,jj}+u_{j,ij}) .\\
\end{aligned} \end{equation}
The boundary conditions over the traction-free surface can be written as,
\begin{equation} \begin{aligned}
    T_j \equiv \sigma_{ij}n_j 
        = 0,\quad {\rm on}\;\partial\Omega .
\end{aligned} \end{equation}
The stress field is generated by the presence of dislocations in the medium, i.e. the ``source terms''.  Formally, these source terms can be written in terms of a discontinuity condition on the displacement field (\citeauthor{hirth&lothe1982}, \citeyear{hirth&lothe1982}),
\begin{equation}
    \oint_{\rm C} \frac{\partial u_j}{\partial l}\, dl = b_j
    \label{eq:burgers_circuit}
\end{equation}
where ${\rm C}$ is any closed path (i.e. the Burgers circuit) that encloses the dislocation line and $b_j$ is the Burgers vector of the dislocation.

The boundary value problems in figure~\ref{fig:decompose_problem}(a) and (b) satisfy the same PDE as in (c) but are subjected to different boundary conditions.
Furthermore, the solutions in (c) is the superposition of the solutions in (a) and (b), and we use superscripts $^{\rm tot}$, $^\infty$, $^{\rm img}$ to designate the solutions in (c), (a), (b), respectively. 
\begin{equation} \begin{aligned}
    u_j^{\rm tot} = u_j^{\infty} + u_j^{\rm img},\quad
    \sigma_{ij}^{\rm tot} = \sigma_{ij}^{\infty} + \sigma_{ij}^{\rm img},
    \quad {\rm in}\;\Omega \\
    T_j^{\rm tot} = T_j^{\infty} + T_j^{\rm img} = 0,\quad
    {\rm on}\;\partial\Omega
\end{aligned} \end{equation}
The problem in figure~\ref{fig:decompose_problem}(a) corresponds to dislocations in an infinite medium.  So the displacement field $u_j^\infty$ satisfies the discontinuity condition, equation~(\ref{eq:burgers_circuit}), which is the only boundary condition for problem (a).
However, the traction field $T_j^\infty$ evaluated on $\partial\Omega$ is generally non-zero.
Because the stress field of a dislocation in an infinite isotropic medium is available analytically (\citeauthor{hirth&lothe1982}, \citeyear{hirth&lothe1982}), $T_j^{\infty}$ can be easily evaluated on $\partial\Omega$.

The problem in figure~\ref{fig:decompose_problem}(b) corresponds to an elastic medium containing a spherical hole but no dislocations.  So its displacement field $u_j^{\rm img}$ is continuous and no longer needs to satisfy the discontinuity condition, equation~(\ref{eq:burgers_circuit}).  The stress source for this problem is the traction boundary condition:
\begin{equation} \begin{aligned}
    T_j^{\rm img} = -T_j^{\infty} \quad {\rm on}\;\partial\Omega .
\end{aligned} \end{equation}







\vspace{0.2in}


\bibliographystyle{model1-num-names}
\bibliography{Spherical_Harmonics}







\end{document}